# A region-specific brain dysfunction underlies cognitive impairment in long COVID brain fog

Jinhao Yang[a,b,1], Shaojiong Zhou[c,1], Zhibin Wang[c], Jiahua Xu[a,b], Jia Cheng[a,b], Zhouqian Yin[a,b], Tao Wei[c], Chaofan Geng[c], Xiaoduo Liu[c], Xiang Li[a,b], Xiaoyu Zhou[a,b], Kun Li[d,e], Ruolei Gu[f,g], Raymond Dolan[a,h,i], Yi Tang[c,j],* & Yunzhe Liu[a,b],*

a. State Key Laboratory of Cognitive Neuroscience and Learning, IDG/McGovern Institute for Brain Research, Beijing Normal University, Beijing, China

b. Chinese Institute for Brain Research, Beijing, China

c. Department of Neurology & Innovation Center for Neurological Disorders, Xuanwu Hospital, Capital Medical University, National Center for Neurological Disorders, Beijing 10053, China

d. Department of Physical Therapy, Shandong Daizhuang Hospital, Jining, Shandong, China

e. Jining Key Laboratory of Neuromodulation, Jining, Shandong, China

f. State Key Laboratory of Cognitive Science and Mental Health, Institute of Psychology, Chinese Academy of Sciences, Beijing 1001101, China

g. Department of Psychology, University of Chinese Academy of Science, Beijing 100049, China

h. Max Planck University College London Centre for Computational Psychiatry and Ageing Research, London, UK

i. Department of Human Neuroimaging, Institute of Neurology, UCL, London, UK

j. Key Laboratory of Neurodegenerative Diseases, Ministry of Education of the People's Republic of China, Beijing 100053, China

[1] Those authors contributed equally to this work.

* Correspondence: tangyi@xwhosp.org (Y. Tang); yunzhe.liu@bnu.edu.cn (Y. Liu).

**Abstract**

Long COVID "brain fog" is a common and debilitating subjective syndrome often associated with persistent cognitive impairment after COVID-19 infection. Here, we identify a specific regional brain dysfunction mediating this impairment and provide evidence that targeted neuromodulation improves the deficit. In 120 patients with long COVID brain fog, we found an aberrant perceptual processing pattern. Patients with more severe brain fog committed significantly more false alarms (impulsive responses to non-signals) despite a preserved rate of correct responses. High-density (128-channel) EEG and structural MRI analyses converged on a right inferior insula deficit, characterized by a blunted evidence-monitoring neural signal and cortical atrophy. We confirmed this finding in a separate 796-participant UK Biobank longitudinal COVID re-imaging cohort. In this cohort, COVID-19 survivors showed selective impairment on a perceptual processing task and corresponding longitudinal atrophy of the right inferior insula compared with healthy controls. Finally, in a proof-of-principle randomized, sham-controlled trial (n = 40), non-invasive excitatory theta-burst ultrasound stimulation targeting the right inferior insula rescued the perceptual deficit by reducing false alarms. These findings provide converging evidence of a causal role for right inferior insula dysfunction in long COVID-related perceptual impairment and show that modulation of this region can rescue the deficit, establishing it as a potential novel therapeutic target.

**Introduction**

Cognitive impairment is among the most frequent and disabling symptoms of long COVID[1,2]. The subjective experience of this impairment is a constellation of symptoms often termed "brain fog," encompassing difficulties with concentration and memory, mental fatigue, and confusion[3,4]. Of significant concern, this condition affects an estimated 10-30% of COVID-19 patients[5], with reports of symptoms persisting for as long as 2.5 years post-infection[6]. Despite the clinical prominence of cognitive impairment, its underlying neural mechanisms remain poorly characterized, posing a major barrier to developing targeted and effective treatments[7] for long COVID patients.

Converging evidence suggests that the neurological sequelae of COVID-19 can cause region-specific brain injury[8-10]. Notably, large-scale MRI studies of COVID-19 survivors have revealed gray matter loss in fronto-limbic areas, including the orbitofrontal cortex, anterior cingulate cortex, parahippocampal gyrus, and insula[11,12]. Although these structural abnormalities are increasingly documented[8,11,12], their functional consequences, especially how they directly account for cognitive impairments, remain unknown. To bridge this gap, it is necessary to move beyond subjective reports and pinpoint a primary cognitive deficit.

The most robust evidence for a specific cognitive deficit in COVID-19 is in the domain of perceptual processing, as indicated by consistently worse performance on the Trail Making Test (TMT), a standard measure of this cognitive function[13]. This deficit appears across disease stages, being observed during the acute illness[14,15] and persisting for months to a year into recovery[16], and it is documented in both hospitalized and non-hospitalized cases[17-20]. Crucially, a longitudinal UK Biobank study revealed that even mild COVID-19 cases show a specific slowing on the TMT approximately five months post-infection, with no concurrent declines in other cognitive domains[11]. While the TMT effectively flags a perceptual processing impairment in long COVID, it does not pinpoint the associated neural dysfunction or specify which precise aspect of perceptual cognition is disrupted. Therefore, a more granular experimental paradigm is needed to link the cognitive impairment to the underlying brain circuitry.

Here we employed the continuous random dot motion (cRDM) task to probe the core mechanisms of perceptual processing. The cRDM paradigm requires participants to monitor sensory evidence from moving dot stimuli to make a direction judgment as coherent motion emerges[21-23]. Combining this precise behavioral task with multimodal neuroimaging and neuromodulation allows us to investigate the neural substrates of impaired perceptual processing and establish a direct link between the cognitive deficit and its underlying brain circuitry.

We conducted a series of three progressive studies (**Fig. 1**). In Study 1 (n = 120), using the cRDM task with concurrent high-density EEG and structural MRI, we characterized perceptual deficits in a cohort of long COVID patients with brain fog. In Study 2, we validated our findings in a large independent sample from the UK Biobank (UKB), examining longitudinal MRI and cognitive data acquired both before and after COVID-19 infection (393 matched case-control pairs; n = 786). Together, these studies provided converging evidence that right inferior insula dysfunction is related to a core cognitive deficit in long COVID brain fog. Notably, the inferior insula is a multisensory, salience-processing region implicated in cognitive monitoring and interoception[24,25], and it has also emerged as a site of structural vulnerability in COVID-19[11].

We further posited that if this region is causally involved in perceptual deficits, then modulating its activity would remediate the cognitive deficit in long COVID brain fog. Thus, in Study 3 (n = 40; 20 sham), we conducted a randomized, sham-controlled proof-of-principle trial of transcranial ultrasound stimulation (TUS). TUS is a non-invasive technique that uses focused ultrasound to precisely modulate deep brain activity[26,27]. We employed an excitatory theta-burst protocol[27-29] targeting the right inferior insula and assessed behavioral changes in perceptual cognition using the cRDM task. This integrated approach allowed us to pinpoint the right inferior insula as a mechanistic and potentially treatable driver of cognitive impairment in those with long COVID.

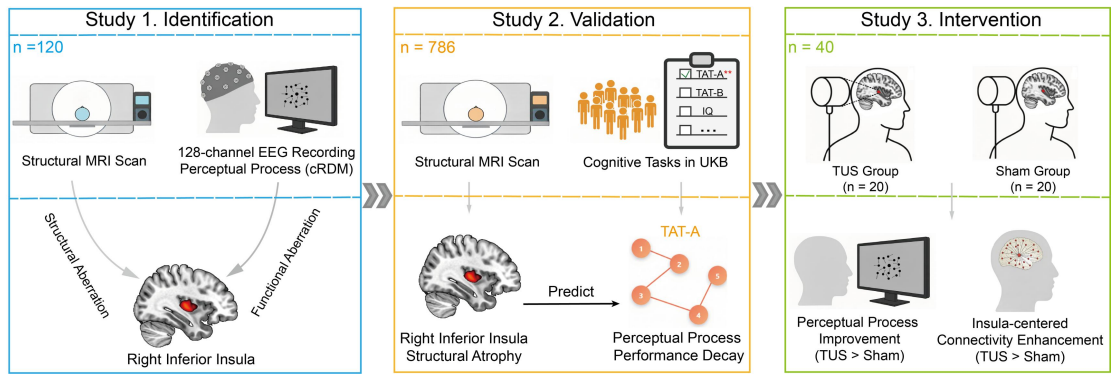

**Fig.1 | Overview of the study design and findings.**

We conducted a progressive, multi-study investigation to identify and causally validate the neural mechanism accounting for long COVID perceptual processing deficits. First, Study 1 characterized perceptual deficits in 120 patients using a cRDM task combined with high-density EEG and structural MRI, pinpointing dysfunction of the right inferior insula as a key neural correlate of impaired performance. Next, Study 2 validated this finding in a large, independent longitudinal dataset from the UK Biobank (UKB; n = 786), where analysis of pre- and post-infection data confirmed that structural abnormalities in the right inferior insula were specifically associated with post-COVID perceptual decline. Finally, to establish a causal link, Study 3 employed a proof-of-principle, sham-controlled, randomized trial (n = 40; 20 sham) of transcranial ultrasound stimulation (TUS) using an excitatory theta-burst protocol targeting the right inferior insula; perceptual cognition was assessed using the cRDM task. Enhancing this region's activity improved perceptual performance (reducing false alarms), establishing its causal role while highlighting a potential therapeutic target.

**Results**

**Brain fog severity correlates with higher false alarm rate during perceptual decisions**

In Study 1, we recruited 120 patients who reported persistent brain fog symptoms at least four weeks after confirmed COVID-19 infection (see Methods for inclusion criteria). All participants were required to complete the cRDM task while undergoing 128-channel EEG recording and undergo a structural brain MRI scan after the task. Participants completed a Brain Fog rating scale[30], alongside demographic and clinical assessments. After excluding 7 individuals with incomplete data or poor-quality EEG, 113 patients remained for analysis (demographics in **Table 1**). Notably, patients who had more severe acute COVID-19 illnesses tended to report more severe brain fog later (Spearman's $\rho$ = 0.411, $p$ < 0.001, controlling for age, sex, and education).

In the task, participants continuously monitored a field of moving dots whose motion coherence fluctuated over time[4,23]. They were instructed to press a button as soon as they perceived continuous leftward or rightward motion of the dots. A correct response was defined as an accurate direction judgment made within a specific time window following the onset of sustained coherent motion, requiring participants to make a reliable judgment based on continuous evidence. In contrast, a button press outside this window, was classified as a false alarm (**Fig. 2a**; see Methods for details), where perceived motion arose not from the true signal, but only from transient, random coherence fluctuations. Notably, although these fluctuations could be large in amplitude, they were not sustained.

To keep participants aware of their performance, all responses were followed by immediate visual feedback (e.g., green for correct, red for a false alarm). We included two conditions that differed only in the duration of "stationary" periods (zero-signal coherence). In the frequent condition, zero-signal coherence intervals lasted 3–8 s, whereas in the rare condition they lasted 5–40 s, making sustained coherent motion more common in the frequent condition (**Fig. 2b**).

We first verified that our cRDM paradigm produced the expected pattern of behavior. Task performance improved with stronger motion signals. Specifically, the correct response rate increased with higher signal coherence (three levels: 0.3, 0.4, 0.5), and was overall higher in the frequent condition than the rare condition (two-way ANOVA, coherence effect: $F(2,224) = 335.72$, $p < 0.001$, $\eta^2 = 0.486$; condition effect: $F(1,112) = 63.49$, $p < 0.001$, $\eta^2 = 0.077$; **Fig. 2c**). Meanwhile, the false alarm rate did not differ between frequent and rare conditions ($t(112) = 0.479$, $p = 0.633$).

To determine whether brain fog severity is linked to impaired perceptual decision-making, we examined correlations between individual brain fog severity and task performance, controlling for age, sex, and education (partial Spearman's correlations). While brain fog severity showed no significant correlation with correct response rate ($\rho = -0.113$, $p = 0.234$), higher brain fog severity was associated with an increased false alarm rate ($\rho = 0.203$, $p = 0.031$; **Fig. 2d**). This relationship held within both task conditions (frequent: $\rho = 0.193$, $p = 0.041$; rare: $\rho = 0.220$, $p = 0.019$), whereas there was no significant relationship between brain fog and correct responses in either condition ($p > 0.2$). Given that the brain fog effect on false alarms was similar in the two conditions, we pooled data across conditions to increase statistical power for subsequent analyses.

**Table 1 | Demographic and clinical characteristics of long COVID patients with brain fog (Study 1)**

|  | n or mean ± s.d. (range) |
|---|---:|
| Number | 113 |
| Age | 34.12 ± 11.61 (18-67) |
| Sex(male/female) | 38/75 |
| Years of education | 16.77 ± 2.69 (9-22) |
| Marital status(married/unmarried/divorced) | 45/67/1 |
| Tobacco smoking(yes/no) | 6/106 |
| Alcohol-intake(yes/no) | 13/99 |
| Comorbidities |  |
|     cardiovascular diseases | 6 |
|     metabolic diseases | 6 |
|     cerebrovascular diseases | 2 |
|     psychiatric or neurological diseases | 5 |
|     allergic diseases | 10 |
| COVID-19 clinical information |  |
|     times of infection(1/2/3) | 90/19/1 |
|     times of vaccinations(0/1/2/3/4) | 6/3/14/84/3 |
|     treated in hospital(yes/no) | 3/107 |
|     acute COVID-19 severity | 10.46 ± 3.90 (0-20) |
|     days since COVID-19 | 163 ± 84 (43-380) |
| Brain fog assessment scale | 17.61 ± 9.38 (0-51) |
| Level of care for brain fog(inpatient/outpatient/home) | 1/13/96 |
| Mini-Mental State Examination(MMSE) | 29.24 ± 1.07 (25-30) |
| Montreal Cognitive Assessment(MoCA) | 27.36 ± 2.28 (19-30) |
| Hamilton Anxiety Rating Scale(HAMA) | 5.80 ± 4.93 (0-31) |
| Hamilton Depression Scale(HAMD) | 4.42 ± 4.11 (0-23) |
| Epworth Sleepiness Scale(ESS) | 8.92 ± 4.21 (0-16) |

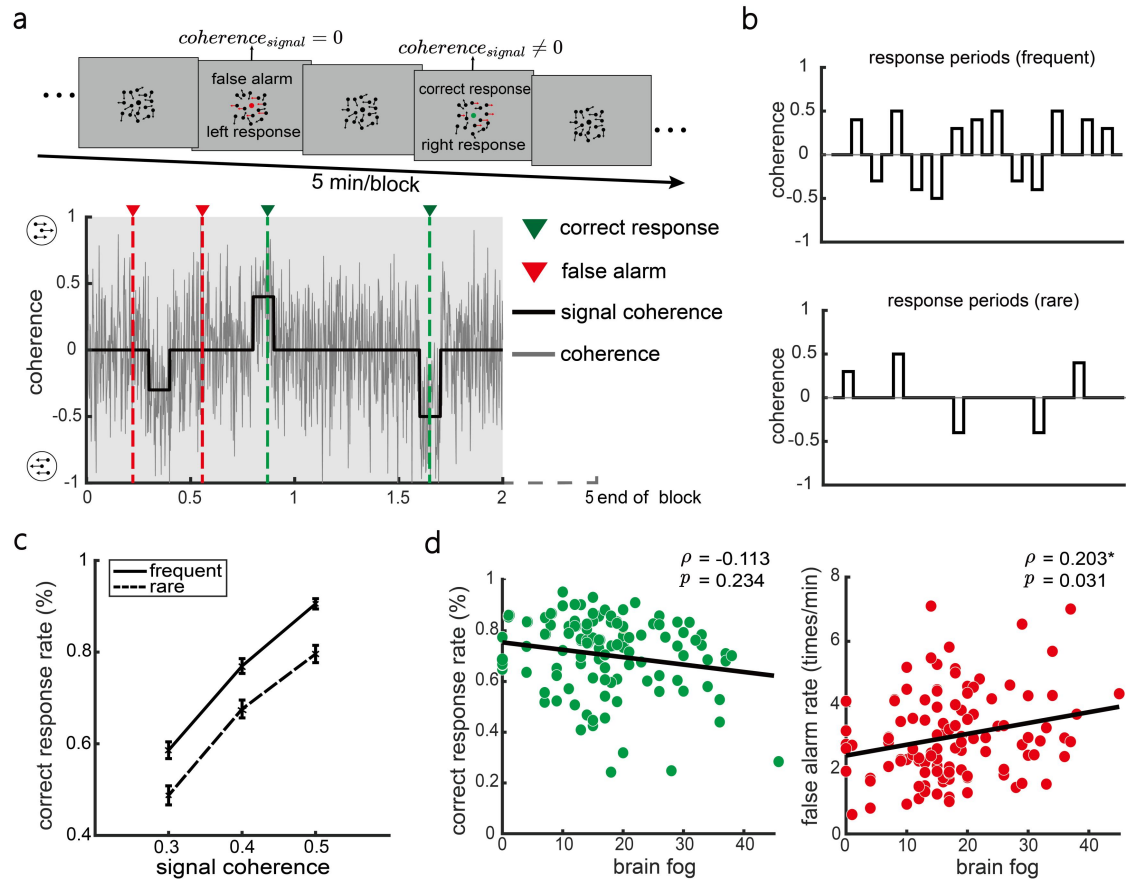

**Fig.2 | Continuous random dot motion (cRDM) task and behavioral performance.**

**a,** Task design. Participants viewed blocks of randomly moving dots whose overall motion coherence (gray line) was determined by adding white noise to a cycling signal (black line). They were instructed to press a left or right key as soon as they perceived continuous motion in that direction (left for coherence < 0, right for coherence > 0, each has three levels).

**b,** Task conditions. In the frequent condition, zero-signal-coherence (no motion) intervals lasted 3-8 s; in the rare condition, zero-signal-coherence intervals lasted 5-40 s. Non-zero signal coherence epochs lasted for a maximum of 5 s, terminating early if a response is made.

**c,** Correct response rate (proportion of sustained coherent motion events correctly identified) increased with the absolute motion signal strength (coherence) in both conditions (N = 113). Error bars denote $\pm 1$ standard error of the mean (s.e.m.).

**d,** Behavior-symptom correlation. Higher brain fog severity was associated with a higher false alarm rate (total false alarms divided by task duration in minutes, red points) but not with correct response rate (green points). Partial correlations that controlled for age, sex, and education are shown. Participants with more severe brain fog committed more false alarms, indicating a greater tendency to respond to transient non-target motion.
x9

**Altered evidence monitoring in individuals with severe brain fog**

The finding that brain fog severity predicts an increased tendency for false alarms (but not a reduced correct response rate) suggests a specific perceptual processing deficit. To detail this, we first compared the time-course of motion coherence preceding false alarms versus correct responses. For each response, we computed an integration kernel[23] by averaging the absolute coherence over the 6-second window before the button press (see Methods), providing a clear and intuitive representation of the stimulus pattern. This revealed a characteristic build-up pattern: on average, motion coherence started near zero and then rose to a peak approximately 0.5 s before a response (**Fig. 3a**, upper panel).

Intriguingly, the coherence traces before false alarms showed a different profile from those seen before correct responses. Just prior to false alarm, motion coherence climbed more steeply and reached a higher peak value than that observed before correct responses. To quantify this, we fit an exponential rise model to each participant's integration kernel[23,31]:

$$k(t) = A e^{\frac{t}{T}}$$

This model estimates two free parameters: $A$, the asymptotic peak coherence, and $T$, the time constant (where a smaller $T$ indicates a faster, steeper rise). The model provided a good fit across participants (average $R^2$ = 0.839). False alarms were preceded by a significantly larger $A$ (indicating a higher peak coherence before a decision) and a significantly smaller $T$ (indicating a faster rise) compared to correct responses (paired t-tests; $A$: $t(112)$ = -8.467, $p < 0.001$; $T$: $t(112)$ = 14.03, $p < 0.001$). In other words, false alarm decisions were triggered by a sharper and higher transient spike in perceived motion evidence (**Fig. 3a**, lower panel).

We next asked whether these integration characteristics were related to reported brain fog severity. For correct responses, brain fog severity did not correlate with either $A$ or $T$ ($A$: $\rho$ = 0.095, $p$ = 0.316; $T$: $\rho$ = -0.096, $p$ = 0.312). For false alarms, brain fog severity showed no significant correlation with the peak $A$ ($\rho$ = 0.167, $p$ = 0.077) but did show a significant correlation with the rise time: individuals with more severe brain fog had smaller $T$ values for false alarms ($\rho$ = -0.305, $p$ = 0.001). Notably, this correlation was significantly stronger than that for correct responses (Steiger's $Z$ = -2.37, $p$ = 0.018). This finding highlights that greater brain fog severity was associated with a steeper coherence ramp leading into false alarm decisions (**Fig. 3b**). In other words, patients with severe brain fog were especially prone to make a decision in response to abrupt coherence surges.

Building on these observations, subsequent analyses aimed to clarify the correlation between brain fog and $T$. Specifically, we sought to test whether this association is underpinned by brain fog compromising monitoring of steadily accumulated evidence over time[21,32], with patients relying more on fleeting spikes in sensory evidence.

To test this formally, we constructed a Generalized Linear Model (GLM) that incorporated regressors for both long-term and short-term evidence to assess their unique contributions to each decision (judging motion vs. no motion). We quantified long-term evidence as the absolute value of cumulative coherence in three successive 0.5 s time bins over the 1.5 s before a response, and short-term evidence as the absolute value of cumulative coherence in fifteen finer-grained 0.1 s bins within the same 1.5 s window (**Fig. 3c**; see Methods for details).

We fitted this GLM separately for correct responses and false alarms. To ensure the stability of the regression estimates, we tested for multicollinearity using Variance Inflation Factors (VIF). For correct responses, VIF values ranged from 2.104 to 5.965 ($VIF_{mean}$ = 3.641), and for false alarms, VIF values ranged from 1.966 to 4.370 ($VIF_{mean}$ = 3.125). All values are well below the threshold of 10[33], indicating that multicollinearity is not a major concern in the analysis.

The GLM results confirmed that sustained evidence is critical for correct decisions. Across participants, higher long-term evidence robustly drove judgments of continuous motion (significant positive weights for all three 0.5 s bins, e.g., 0-0.5 s bin: $t(112)$ = 22.264, $p$ <

0.001), and long-term evidence predictors were much more influential for correct responses than for false alarms, especially in the final 0-0.5 s before a response ($t(112)$ = 11.774, $p < 0.001$; **Fig. 3d**, left).

In contrast, false alarm decisions were disproportionately triggered by last-moment noise, as shown by the finding that only in false alarms did the immediate 0.1 s prior to a response show a positive coherence effect ($t(112)$ = 8.172, $p < 0.001$), whereas for correct responses this final 0.1 s had a negative weight ($t(112)$ = -7.940, $p < 0.001$), reflecting that a sudden random perturbation often interrupted a correct decision.

The weight on this final 0.1 s bin was significantly larger (more excitatory) for false alarms than for correct responses ($t(112)$ = -11.045, $p < 0.001$; **Fig. 3e**, left). Critically, brain fog severity correlated with these effects: patients with more severe brain fog showed a diminished monitoring of long-term evidence (the weight on the -0.5-0s bin for false alarms correlated negatively with brain fog severity, $\rho = -0.237$, $p = 0.012$; **Fig. 3d**, right) and a heightened reliance on last-instant spikes (the weight on the -0.1-0 s bin in false alarms correlated positively with brain fog severity, $\rho = 0.283$, $p = 0.003$; **Fig. 3e**, right). Thus, as brain fog severity increases, perceptual decision-making becomes increasingly driven by momentary noise and less guided by sustained evidence. This manifests as an increased false alarm rate.

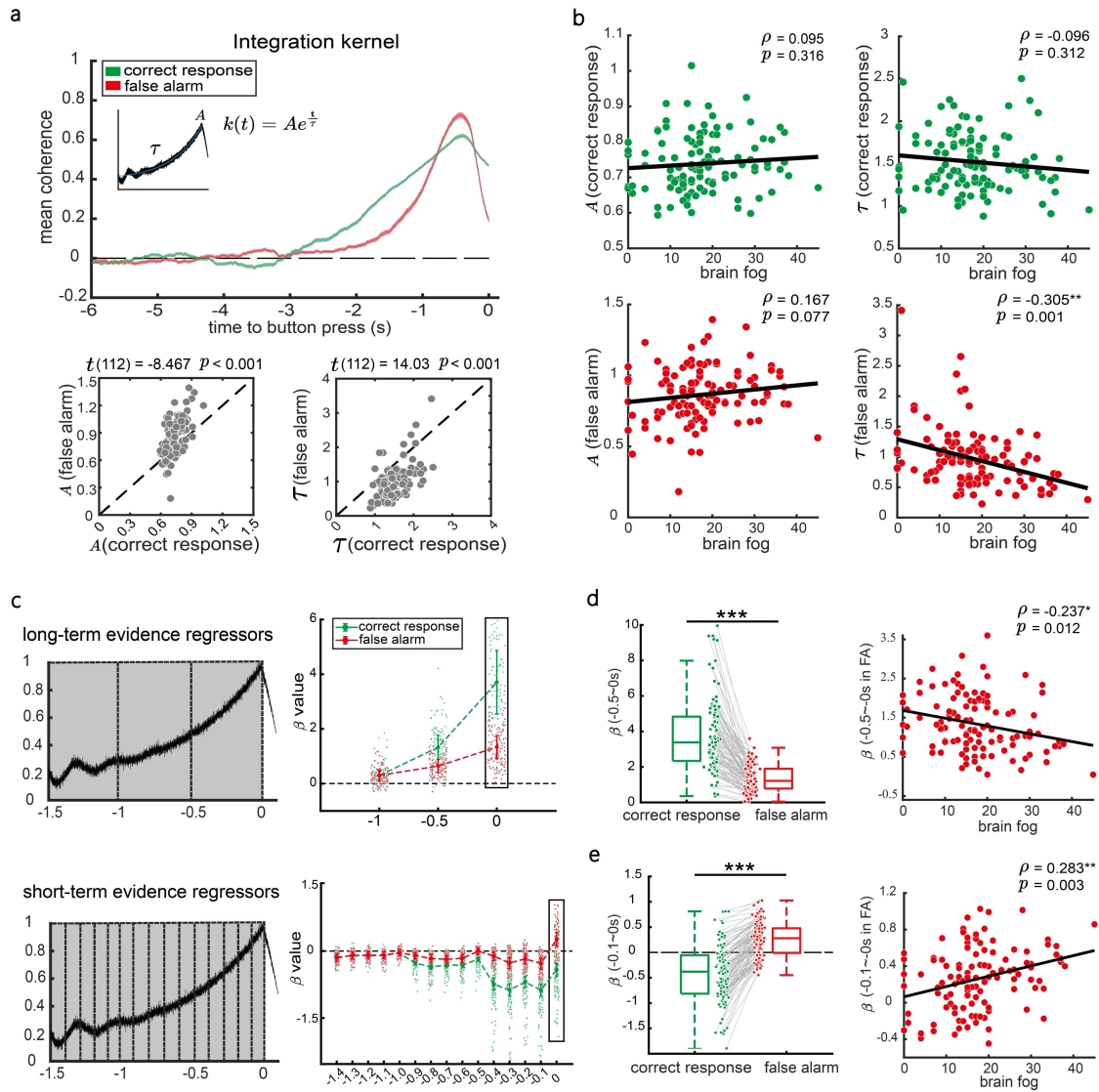

**Fig.3 | Coherence patterns reveal altered evidence monitoring in brain fog.**

**a,** Integration kernel analysis. We averaged the stimulus coherence in the 6 s leading up to each response (time 0 = response). On average, coherence rose to a peak ~0.5 s before a response. This rise was fit by an exponential model with parameters: $A$ (peak coherence) and $\tau$ (rise time constant). False alarms (red) showed a higher peak $A$ and a steeper rise (smaller $\tau$) than correct responses (green).

**b,** Parameter-symptom correlation. Partial spearman's correlations (controlling for age, sex, and education) revealed no significant links between brain fog severity and $A$ or $\tau$ for correct responses (green). For false alarms (red), brain fog severity did not correlate with $A$, but showed a negative correlation with $\tau$, meaning more severe brain fog was associated with a faster (steeper) coherence ramp prior to false alarms.

**c,** GLM with long-term and short-term evidence. We constructed a GLM to predict motion perception decisions. long-term evidence regressors (upper panel) captured the magnitude of accumulated coherence in three 0.5 s bins over the final 1.5 s before a response. short-term evidence regressors (lower panel) captured rapid coherence changes in fifteen 0.1 s bins in the same window. In general, stronger long-term evidence increased the probability of perceiving motion (for both correct and false-alarm trials), especially closer to the decision. However, in false alarm trials, only the last 0.1 s before the response showed a positive short-term coherence effect.

**d,** Monitoring of long-term evidence. The cumulative evidence in the final 0-0.5 s bin was significantly more predictive of correct responses than of false alarms. Furthermore, brain fog severity correlated negatively with the false alarm weight for this bin, indicating that individuals with severe brain fog relied less on sustained evidence prior to false alarms.

**e,** Susceptibility to momentary spikes. The short-term evidence regressor for the last 0.1 s (-0.1-0 s) was significantly larger (more positive) for false alarms than for correct responses, reflecting false alarm decisions being driven by abrupt coherence spikes that were successfully ignored during correct responses. This last-moment weight also correlated positively with brain fog severity, indicating patients with more severe brain fog were more prone to false alarms caused by fleeting noise fluctuations.

# Right inferior insula dysfunction as a critical neural substrate of perceptual deficit in brain fog

Having identified the perceptual deficit in brain fog (an over-reliance on brief evidence spikes and failure of sustained monitoring), we next investigated the neural mechanisms underlying this impairment. We analyzed the 128-channel EEG recordings time-locked to task events, using a deconvolutional GLM to separate overlapping event-related signals[23,34]. This enabled us to isolate neural activity associated with specific events (stimulus changes, responses, outcomes) across time (**Fig. 4a**, left; see Methods for more details).

Differences between correct and false alarm trials emerged around the response and feedback. In false alarm trials, we observed a distinctive pattern of EEG activity over centroparietal electrodes (E55, E37, E87, E30, E105, E129, E7, E31, E106, E80; **Fig. 4a**, right). Approximately 1 s before a false alarm response, the centroparietal signal (averaged over centroparietal electrodes) showed a significant suppression relative to baseline. Immediately following a button press, this was followed by a rapid positive surge, then a dip, and then a second rebound positive peak ~0.4-0.6 s after the response (**Fig. 4a**, right; see also **Extended Data Fig. 1**, showing the full time courses of "correct response" and other regressors). We refer to this late rebound positivity (0.4-0.6 s post-response) as the post-response centroparietal positivity (PCPP).

The PCPP was most closely tied to brain fog deficits compared with other time windows (**Extended Data Fig. 2**). Participants with a smaller (attenuated) PCPP tended to have a steeper rise in evidence integration (smaller $\Tau$) and higher brain fog severity. Specifically, PCPP amplitude at 0.4-0.6 s post-response correlated positively with the false-alarm integration parameter $\Tau$ ($\rho$ = 0.225, $p$ = 0.017) and negatively with brain fog severity ($\rho$ = -0.297, $p$ = 0.001), after controlling for covariates. The correlation with brain fog severity was the strongest observed among all EEG features (see also **Extended Data Fig. 2**), suggesting that reduction of the PCPP is a core neural marker of brain fog severity.

Consistent with this, mediation analysis indicated that the PCPP partly mediated the relationship between brain fog severity and the false alarm rise parameter $T$. Lower PCPP amplitude contributed to steeper coherence ramp prior to false alarms in those with more severe brain fog (**Fig. 4b**; bootstrap test; a: −0.008 ± 0.003, 95% CI = [-0.014, -0.002]; b: 0.3 ± 0.144, 95% CI = [0.015, 0.584]; a×b: −0.002 ± 0.027, 95% CI = [-0.11, -0.001]; c: −0.018 ± 0.005, 95% CI = [-0.027, -0.009]; c' = −0.015 ± 0.005, 95% CI = [-0.025, -0.006]).

This link between PCPP reduction and brain fog led us to ask which brain regions generate the PCPP and whether they are structurally compromised in long COVID. The 128-channel whole-brain EEG recordings and individual structural MRI scans allowed us to address this question. Using sLORETA source localization[35] focused on the PCPP time window, we found that the attenuated PCPP originated primarily in the right inferior insula and the left mid-cingulate cortex (**Fig. 4c**; voxel-wise family-wise error (FWE) corrected, based on Bonferroni method, p < 0.001; see also **Extended Data Fig. 3** for results of other typical electrophysiological potentials). These regions are known to play pivotal roles in salience detection, performance monitoring, and allocation of cognitive resources[24,25,36].

In the same cohort, we then performed a surface-based morphometry (SBM) analysis to examine whether structural features in these source-localized regions were associated with the increased false alarm rate. We focused on cortical thickness given its established sensitivity to cognitive impairment[37,38]. This analysis revealed a significant negative correlation in right inferior insula, which showed the strongest association. Specifically, patients with thinner cortex in this region exhibited a significantly higher false alarm rate ($\rho$ = -0.219, *p* = 0.021; **Fig. 4d**). In contrast, the correlations for the anterior ($\rho$ = -0.182, *p* = 0.056) and posterior middle cingulate cortex ($\rho$ = -0.136, *p* = 0.156) did not reach significance. These findings suggest that reduced cortical thickness in the right inferior insula, potentially reflecting post-COVID atrophy, is linked to a greater tendency toward impulsive responses during perceptual decision-making. They also identify the right inferior insula as a critical neural substrate of long COVID brain fog.

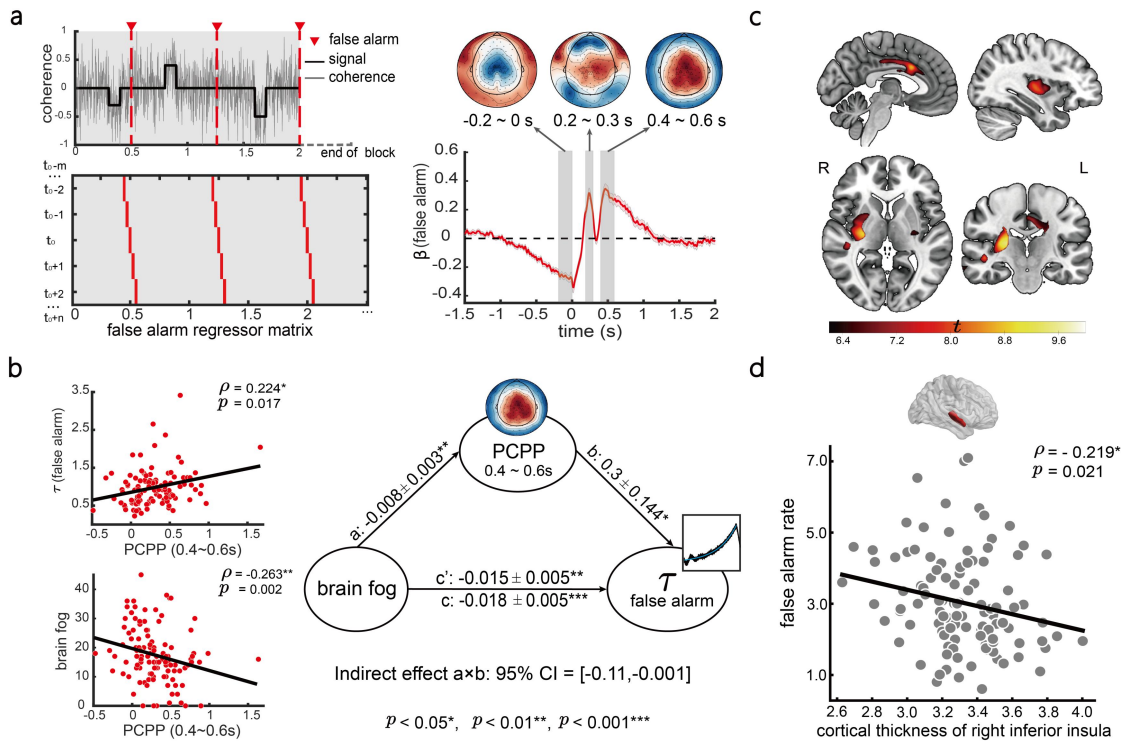

**Fig.4 | Structural and functional deficits in the right inferior insula linked to brain fog.**

**a,** EEG dynamics for false alarms. Using deconvolutional GLM, we isolated neural activity around the time of false alarm responses. Left: Schematic of the analysis aligning EEG to events. Right: In centroparietal electrodes, activity dropped below baseline ~1 s before a false alarm, then after the response showed a sharp positive surge, a dip, and a rebound positive peak. We define this rebound (0.4-0.6 s post-response) as the post-response centroparietal positivity (PCPP).

**b,** Symptom and behavior relationship. PCPP amplitude negatively correlated with brain fog severity and positively correlated with the false alarm rise parameter $\tau$. Mediation analysis indicated that a reduced PCPP partially mediated the link between high brain fog severity and steeper coherence ramp prior to false alarms (parameter $\tau$).

**c,** Source localization of PCPP. sLORETA identified significant generators of the PCPP in the right inferior insula and left middle cingulate cortex (voxel-wise FWE corrected, $p < 0.001$).

**d,** Structure-behavior relationship. SBM analysis showed that the cortical thickness of the right inferior insula correlated with false alarm rate, where patients with greater thinning in this region made more false alarm responses.

**Longitudinal cohort links right inferior insula atrophy to post-COVID perceptual decline**

Thus far, our findings from Study 1 identified the right inferior insula as a candidate region whose dysfunction contributes to the perceptual deficit in brain fog, based on detailed EEG and MRI evidence in a dedicated patient cohort. In Study 2, we sought to cross-validate this region's involvement in an independent, large-scale dataset: the UKB COVID-19 re-imaging study. This rich longitudinal dataset allowed us to test whether COVID-related cognitive changes and brain changes in the general population align with the specific deficit and neural substrate we pinpointed in Study 1.

We assembled Study 2 sample from the UKB longitudinal cohort who underwent brain MRI both before and after the onset of the COVID-19 pandemic. From this cohort, we identified 1,085 individuals who tested positive for SARS-CoV-2 (cases) and 1,011 demographically matched individuals with no infection (controls). To isolate the effects of COVID-19, we excluded any participants with major neurological or psychiatric conditions known to affect cognition (e.g., dementia, schizophrenia, stroke, etc.), while also requiring that participants had completed the full set of cognitive tests (see the Methods for details) at both imaging time points (pre- and post-pandemic). After these exclusions, 687 cases and 630 controls remained. We then applied Mahalanobis distance matching[39] on baseline cognitive test scores to ensure that the selected case and control groups were well-matched in their pre-COVID cognitive performance (accounting for the multivariate cognitive profile; see Methods and also **Extended Data Fig. 4**). This procedure yielded 393 matched case–control pairs for further analysis (**Fig. 5a**).

Within this longitudinal sample, we examined changes in a battery of ten cognitive metrics (from six different tasks; see the Methods for details) spanning domains that included perceptual processing, memory, executive function, and reasoning (selected based on sensitivity to cognitive impairment by Douaud et al.[11]). We then asked whether post-COVID cognitive changes differed between cases and controls. Strikingly, only one measure showed a significant group difference: the Trail Making Test Part-A (TMT-A) completion time. COVID-19 cases had a larger increase (worsening) in TMT-A time from baseline to follow-up compared to controls (mean change difference = +1.06 seconds, *t*(784) = 2.727, *p* = 0.007, Cohen's *d* = 0.195; **Fig. 5b**). No other cognitive metric showed a significant case – control difference after correction. This finding corroborates earlier reports that visual perceptual processing (as probed by TMT-A) is selectively impaired following COVID-19 infection[11]. In our matched cohort, the specific deficit manifested even though the cases had mostly mild illness, reinforcing this is a robust and unique cognitive consequence of COVID-19.

We next tested whether longitudinal brain changes were associated with this cognitive deficit, focusing on cortical regions highlighted by Study 1. For each participant, we computed the change in cortical thickness between the pre-COVID and post-COVID scans. We then performed a GLM analysis using changes in cortical thickness from baseline to follow-up to predict the corresponding change in TMT-A performance. The right inferior insula, as defined in Study 1, again emerged as the significant predictor in COVID-19 cases: greater cortical thinning in the right inferior insula was associated with a larger increase in TMT-A completion time (i.e. greater slowing) (*β* = −0.21, *p* = 0.029; **Fig. 5c**). Thus, COVID-related atrophy of the right inferior insula specifically tracked the degree of post-COVID perceptual slowing. This finding aligns with our patient cohort findings, underscoring the functional relevance of right inferior insula damage: the critical structural change that predicts the hallmark cognitive impairment of long COVID.

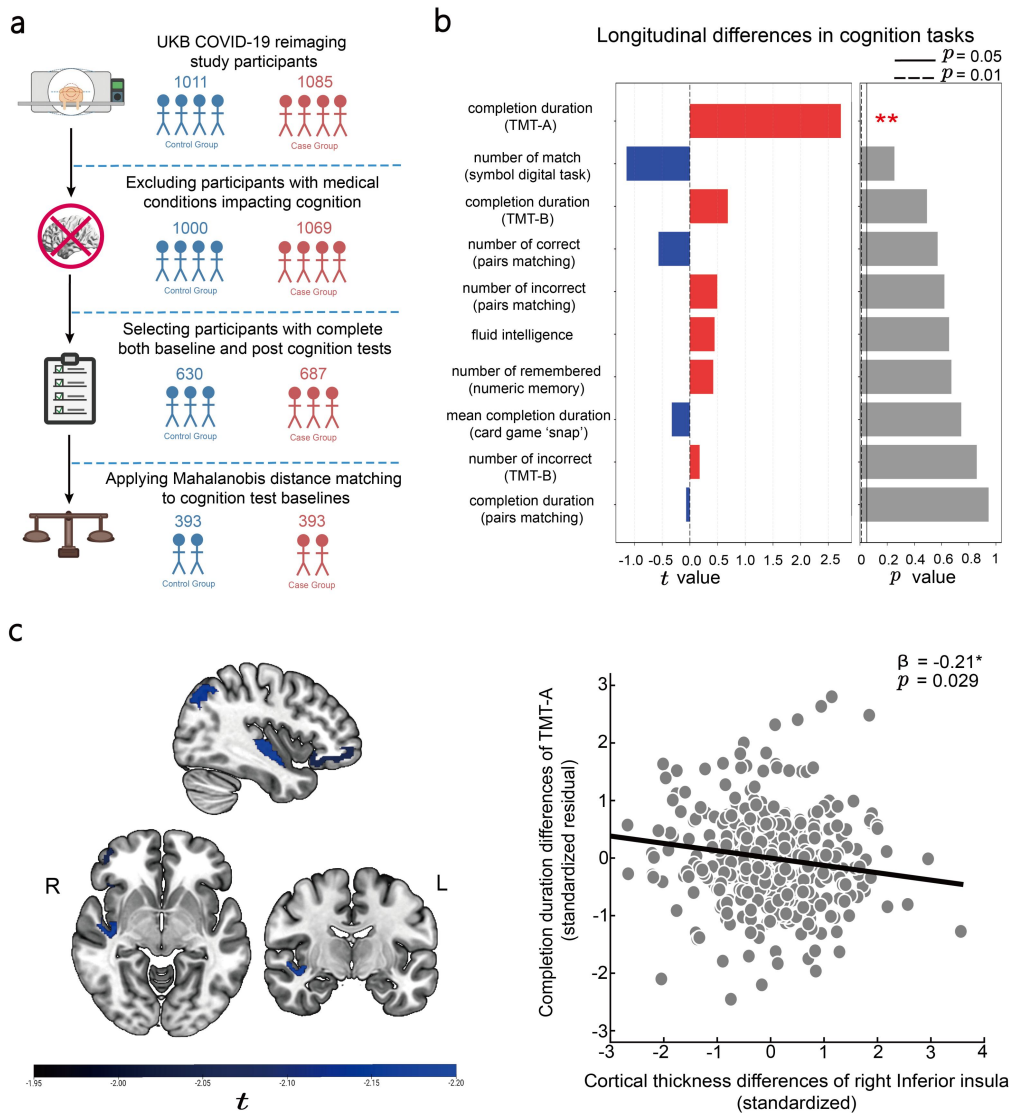

**Fig.5 | Right inferior insula structural aberrations predict COVID-related perceptual deficit (UKB cohort).**

**a,** Cohort selection and matching. From the UKB COVID-19 re-imaging study, we excluded individuals with major neurological conditions or missing data, then matched COVID-19 cases and controls on baseline cognitive performance using Mahalanobis distance. This yielded 393 well-matched case-control pairs.

**b,** Longitudinal cognitive changes. Among ten cognitive metrics examined, only TMT-A showed a significant group difference over time. The COVID-19 cases had a larger increase in completion time (worse performance) compared to controls.

**c,** Brain structure – function link. Greater cortical thinning of the right inferior insula predicted a larger slowing in TMT-A time in COVID-19 cases. In other words, individuals who showed more atrophy in right inferior insula after COVID tended to have a greater decline in visual perceptual processing speed.

**Targeted insula neuromodulation improves perceptual processing in brain fog**

Studies 1 and 2 established a consistent correlation between right inferior insula dysfunction and impaired perceptual cognition in long COVID. To move beyond this correlation and establish causality, Study 3 employed a proof-of-principle, sham-controlled, randomized TUS trial (ID: NCT07154199) to directly modulate the right inferior insula. This intervention aimed not only to confirm the region's causal role in the perceptual deficit but also to explore its potential as a therapeutic target.

We recruited 40 community-dwelling participants with persistent self-reported brain-fog symptoms following COVID-19. Participants were randomly assigned to either the TUS group (n = 20) or the sham group (n = 20), with groups matched on demographics and baseline characteristics (**Table 2**). All participants underwent baseline assessments including a brain-fog severity rating, structural MRI, resting-state fMRI, and the cRDM behavioral task. The TUS group then received low-intensity theta-burst ultrasound stimulation (with excitatory effect[28,29,40]; see Methods for details) targeting the right inferior insula, whereas the sham group underwent a protocol that preserved all auditory and procedural features but emitted no ultrasound. After the intervention, participants repeated the same set of assessments (**Fig. 6a**). Stimulation targeting for each participant was guided by their structural MRI. The target region, the right inferior insula (inferior circular sulcus) defined in the Destrieux atlas[41], was selected to align with structural evidence from Studies 1, 2 and the source-localized functional cluster identified in Study 1. Acoustic simulations were used to verify target pressure and ensure safety (example shown in **Fig. 6b**; see Methods for details).

We first validated the neural effect of our TUS protocol. Using resting-state fMRI, we performed a seed-based functional-connectivity (FC) analysis using the stimulation target in the right inferior insula as the seed. A linear mixed-effects (LME) model revealed a significant overall increase in insula-centered FC after TUS compared to sham ($\beta = 0.073$ 0.028, $p = 0.008$), indicating effective engagement of the target network. The enhanced connectivity was primarily observed with the right frontal pole, superior temporal gyrus, and anterior occipital cortex (see also **Extended Data Fig. 6**).

We next examined whether stimulating the right inferior insula produced behavioral improvements on the cRDM task. There was no significant Group × Time interaction for correct response rate ($F(1,38) = 3.332$, $p = 0.076$; **Fig. 6c** left), and the correct response remained high in both groups. By contrast, a two-way repeated-measures ANOVA of Group (TUS vs. Sham) × Time (Baseline vs. Post) revealed a significant interaction for false-alarm rate ($F(1,38) = 5.800$, $p = 0.021$, $\eta^2 = 0.019$; **Fig. 6c** right), reflecting a specific reduction in false alarms in the TUS group relative to sham. An independent-samples *t*-test showed that the reduction in false alarm rate was significantly greater in the TUS group (mean change: -47.4%) than in the sham group (mean change: -23.7%; $t(38) = 2.408$, $p = 0.021$, Cohen's $d = 0.338$). Immediately after this single stimulation, brain fog symptom ratings did not change significantly ($p = 0.372$). These results indicate that modulating the right inferior insula causally enhanced perceptual performance in long COVID brain-fog patients by specifically reducing impulsive false-alarm errors.

Table 2 | Demographic and clinical characteristics of participants in TUS trial, separately for sham and TUS group (Study 3)

| TUS Group | n or mean ± s.d. (range) |
|---|---|
| Number | 20 |
| Age | 25.25 ± 5.41 (19-39) |
| Sex(male/female) | 9/11 |
| Years of education | 15.40 ± 0.88 (14-16) |
| Marital status(married/unmarried/divorced) | 4/16/0 |
| Tobacco smoking(yes/no) | 3/17 |
| Alcohol-intake(yes/no) | 1/19 |
| Comorbidities | |
|     cardiovascular diseases | 1 |
|     metabolic diseases | 0 |
|     cerebrovascular diseases | 0 |
|     psychiatric or neurological diseases | 0 |
|     allergic diseases | 0 |
| Days since COVID-19 | 713.3 ± 253.4(154-1010) |
| Brain fog assessment scale | 15.05 ± 8.09 (1-39) |

| Sham Group | n or mean ± s.d. (range) |
|---|---|
| Number | 20 |
| Age | 27.75 ± 10.10 (20-57) |
| Sex(male/female) | 4/18 |
| Years of education | 16.05 ± 2.01 (12-19) |
| Marital status(married/unmarried/divorced) | 6/14/0 |
| Tobacco smoking(yes/no) | 1/19 |
| Alcohol-intake(yes/no) | 3/17 |
| Comorbidities | |
|     cardiovascular diseases | 1 |
|     metabolic diseases | 1 |
|     cerebrovascular diseases | 0 |
|     psychiatric or neurological diseases | 0 |
|     allergic diseases | 0 |
| Days since COVID-19 | 719.4 ± 262.3(99-964) |
| Brain fog assessment scale | 15.35 ± 7.62 (3-32) |

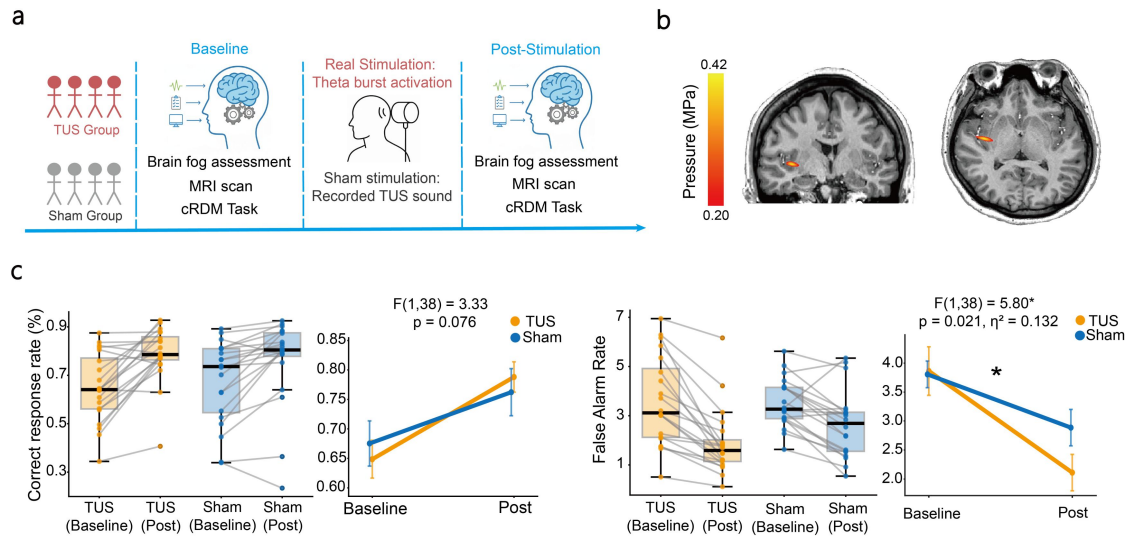

**Fig.6 | Inferior insula-targeted TUS improves perceptual cognition.**

**a,** Study design. Participants with brain fog were randomly assigned to TUS or Sham groups. All underwent baseline assessments (brain fog rating, MRI, resting-state fMRI, cRDM task), then received either theta-burst TUS or sham (sound-only) stimulation, followed by post-stimulation assessments.

**b,** Acoustic simulation. The TUS was directed at the right inferior insula (target shown on an example MRI). An acoustic simulation overlay confirms focal targeting; warmer colors indicate higher ultrasound pressure.

**c,** Behavioral outcomes. cRDM task performance before and after intervention for TUS vs. Sham groups. Box plots (left) show correct response rate (top) and false alarm rate (bottom) at baseline and post-stimulation. On the right, interaction plots illustrate the Group × Time effects. False alarm rate showed a significant interaction: the TUS group (orange) improved (decreased false alarms) relative to Sham (blue), whereas correct response did not significantly differ by group.

**Discussion**

We conducted three synergistic studies to map the neural underpinnings of cognitive impairment in long COVID brain fog. We provide evidence for a specific cognitive deficit in sustained evidence monitoring during perceptual decision-making, expressed as an increased tendency to make arbitrary and impulsive judgments driven by transient sensory noise rather than accumulated evidence. We identify both functional and structural abnormalities in the right inferior insula as a core neural substrate of this deficit and show that the perceptual deficit can be rescued by targeted neuromodulation. By integrating mechanistic investigation (Study 1), longitudinal validation (Study 2), and targeted neuromodulation (Study 3), our work provides robust evidence for a causal role of the right inferior insula in long COVID cognitive impairment and points to a promising therapeutic target.

Our behavioral results offer a nuanced view of this core deficit. Rather than a global decline in perceptual performance, patients with more severe brain fog showed preserved correct responses, but an elevated rate of false alarms. Computational modeling indicated that this arises from impaired monitoring of sustained evidence and heightened influence of transient spikes of sensory input. Effective perceptual decision-making ordinarily requires steadily accumulated evidence over time to form a stable judgment while filtering out momentary noise[42,43]. Brain fog appears to erode this stability, leading to noise-driven perceptual decisions that aligns with patient reports of reduced mental clarity, difficulty sustaining attention, and feeling overwhelmed by sensory stimuli[44].

Our electrophysiological results show that in individuals with more severe brain fog, a reduced PCPP mediated their shift away from sustained evidence and toward transient sensory input, which reflects a deficit in evidence monitoring. Two lines of evidence converge to pinpoint the right inferior insula as the source of this monitoring deficit. First, EEG source localization identified the right inferior insula (along with the left middle cingulate cortex) as the origin of the diminished PCPP signal. Second, structural MRI showed that patients with greater cortical thinning in the right inferior insula had more false alarm responses. This convergence is compelling given the established role of the insula, particularly in the right hemisphere[45], as a hub for integrating internal and external information[24,46,47] and for monitoring task performance and errors[48-50]. Consistent with this role, our results indicate that damage to the right inferior insula impairs the cognitive ability to monitor error signals during decision making, thereby weakening sustained evidence monitoring and permitting noise-driven responses.

More broadly, dysfunction of the right inferior insula provides a unifying explanation for the diverse clinical features observed in post-COVID conditions. The insula has emerged as a consistent site of COVID-related brain changes. Large-scale imaging studies have repeatedly identified the insula among regions with post-COVID structural atrophy[11,12]. The insula is also a key interface for processing the peripheral inflammation thought to drive long COVID pathology[9,10]. It maintains dense bidirectional connections with the autonomic and immune systems, such that insular dysfunction can both result from and contribute to systemic inflammatory processes[51,52]. Moreover, as a primary cortical hub for olfaction and gustation[53], insula provides a direct neural substrate for the loss of smell and taste, which are hallmark symptoms of acute COVID-19[54,55]. We also note that insula abnormalities are not unique to COVID-19. Similar "brain fog" symptoms in other conditions, such as myalgic encephalomyelitis/chronic fatigue syndrome (ME/CFS), have been linked to insular hypoactivation and connectivity changes[56]. These observations suggest that right inferior insula dysfunction may represent a common neurobiological mechanism connecting sensory and inflammatory disruption with cognitive impairment.

TUS permits focal modulation of deep cortical and subcortical structures that are not easily accessible with conventional non-invasive methods[27,57]. It has been shown to alter neurochemistry and functional connectivity in humans[40], and to modulate behavior in primates and humans[26,57], demonstrating promise for clinical translation in neurological disorders[58]. Building on this literature, our trial demonstrates that targeting the right inferior insula with theta-burst TUS[28,29,40] reduces the maladaptive false alarms that characterize long COVID cognitive impairment. The insula is a challenging target due to its depth and gyral anatomy. Nonetheless, our results indicate that this region is amenable to precise, MRI-guided TUS, with measurable behavioral and network effects. This positions TUS as a feasible tool for therapeutic interventions.

While the findings from a single TUS session are encouraging, we did not observe any immediate change in self-reported brain fog symptoms after the intervention in our study. This may reflect the limitations of subjective, retrospective symptom ratings in detecting acute effects. Alternatively, a single dose may be insufficient to produce a durable neural response capable of shifting persistent symptoms[57]. Follow-up studies are warranted to assess the long-term efficacy of repeated TUS sessions[59] and to optimize dosing schedules, paying particular attention to the impact on experiential symptoms. It is also plausible that combining TUS with cognitive training[60] may yield more robust and lasting clinical benefits. The current findings support a translational path for symptom-directed therapy. In this approach, symptom-relevant readouts (false alarms on a perceptual task) serve as proximal endpoints for dose optimization and target engagement, while resting-state connectivity changes provide mechanistic biomarkers of network-level response.

In summary, our findings establish a critical link between right inferior insula dysfunction and impaired perceptual processing in long COVID cognitive deficits. Targeted neuromodulation of the right inferior insula restores perceptual performance, providing a strong rationale for developing insula-based therapies for the cognitive sequelae of COVID-19.

**Methods**

**Participants**

In Study 1, a total of 120 long COVID patients were recruited from communities and assessed at XuanWu Hospital, Capital Medical University between January and October 2023. Following the formal definition of UK National institute for Health and Clinical Excellence[61], patients should have a history of COVID-19 infection (positive test results for either polymerase chain reaction or rapid antigen test) and report persistent brain fog symptoms at least four weeks after COVID-19 infection. The exclusion criteria included: 1.Presence of any symptoms of cognitive impairment or other neurological symptoms prior to COVID-19 infection; 2.Structural MRI revealing significant intracranial lesions or structural abnormalities; 3.Development of severe neurological complications after COVID-19 infection, including delirium, cerebrovascular diseases, encephalitis, and epilepsy; 4. Other disorders that may cause cognitive impairment, including Dementia, Schizophrenia spectrum disorders, stroke, Alzheimer's disease, and Parkinson's disease. After excluding 7 patients due to early discontinuation of the experiment and poor-quality EEG, we had a total of 113 patients for primary analysis.

In Study 2, participants were selected from the UKB COVID-19 Reimaging Study. This study was designed to investigate brain structural aberrations caused by SARS-CoV-2 infection. It invited individuals who had undergone an initial MRI scan prior to the COVID-19 pandemic with no detectable structural abnormalities identified at that time to return for a follow-up MRI scan. Cases were defined as participants with documented positive SARS-CoV-2 antigen or PCR test results, medical records confirming COVID-19 diagnosis, or antibody-positive results in field testing followed by a confirmed retest for unvaccinated individuals; controls were participants with no documented COVID-19 medical history and negative antibody test results, individually matched to cases on sex, race/ethnicity, age ($\pm 6$ months), initial scan center location, and scan date ($\pm 6$ months).

Then, we excluded participants if they had ICD-10 diagnoses that may cause cognitive deficit, including dementia (F00-F07), schizophrenia spectrum disorders (F20-F29), epilepsy (G40-G41), stroke (I60-I64), Alzheimer's disease (G30-G32), Parkinson's disease (G20-G26), and traumatic brain injury (S07). Additionally, only participants who completed both baseline and follow-up cognitive tests were included. Ten cognitive metrics (derived from 6 distinct tasks: TMT, Symbol Digital Task, Pair Matching Task, Fluid Intelligence, Numeric Memory Test, Card Game "Snap") were selected based on their demonstrated sensitivity to cognitive impairment, established by comparing the baseline performance of individuals who later developed dementia with that of matched controls[11]. Given significant differences in baseline fluid intelligence between groups and the presence of intercorrelations and heterogeneous scales among cognitive metrics, Mahalanobis Distance Matching[39], was employed to control baseline cognitive performance, as this method standardizes multidimensional data and corrects covariance dependencies using the inverse covariance matrix[62].

In Study 3, we recruited 40 community participants with persistent brain fog symptoms and a history of COVID-19 infection. Exclusion criteria included: 1. Significant intracranial lesions or structural abnormalities identified via structural MRI; 2. Disorders potentially causing cognitive deficit, such as dementia, schizophrenia spectrum disorders, stroke, Alzheimer's disease, and Parkinson's disease. Participants were randomly assigned to TUS or sham stimulation.

Study 1 and 3 were approved by the ethics review committee of Xuanwu Hospital, Capital Medical University (KS2022105-1). Study 2 utilized data from the UKB, which was approved by the Northwest Multi-Centre Research Ethics Committee (MREC). Study 1 and 3 were also collectively registered on the ClinicalTrials.gov (ID: NCT07154199). Informed consent was obtained from participants in Study 1 and 3 before the experiments began.

**Experimental procedure**

In Study 1, on the day of participation, all patients' brain fog symptom severity was assessed using the Brain Fog Assessment Scale[30]. This 17-item scale evaluates distinct brain fog symptoms, each rated on a 4-point Likert scale (0 = no symptom, 1 = mild, 2 =

moderate, 3 = severe). Participants also completed the cRDM task during 128-channel EEG recording and underwent structural MRI scans. Additionally, the following data were collected: demographic information, COVID-19 clinical history, Mini-Mental State Examination (MMSE)[63], Montreal Cognitive Assessment (MoCA)[64], Hamilton Anxiety Rating Scale (HAMA)[65], Hamilton Depression Rating Scale (HAMD)[65], and Epworth Sleepiness Scale (ESS)[66]. It's important to note that since China lifted its pandemic containment measures in December 2022, the presence of asymptomatic infections has complicated the recruitment of healthy controls. We also did not include a control group of participants with prior COVID-19 infection but no brain fog symptoms, as COVID-induced neurological and cognitive symptoms may not be subjectively recognized.

Study 3 spanned 2 days. On Day 1, participants first completed assessments of their persistent brain fog symptoms using the Brain Fog Assessment Scale and provided additional demographic information. Next, they performed the baseline cRDM task; 20 minutes after completing this task, structural MRI and baseline resting-state MRI data were collected. On Day 2, participants first received 60 seconds of stimulation—either TUS or sham, depending on their randomly assigned group. Twenty minutes after the stimulation, a follow-up resting-state MRI scan (8 minutes) was conducted. Immediately after completing this scan, participants performed the follow-up cRDM task.

**Continuous random dot motion (cRDM) task**

Our cRDM task, adapted from prior research[4,23], required participants to continuously monitor a dynamic stream of dots moving randomly on a screen. The directional consistency of the dots was quantified by a coherence parameter, ranging from -1 to 1. A negative coherence value indicated a leftward motion (more dots moving left), a positive value indicated a rightward motion (more dots moving right), and zero indicated purely random motion with no directional bias. For calculation, the coherence value combined signal coherence with added white noise. Signal coherence comprised three levels (0.3, 0.4, 0.5). To account for motion direction, these were assigned positive or negative signs (±0.3, ±0.4, ±0.5), with 0 representing the stationary state. White noise samples were drawn from a Gaussian distribution with a mean of 0 and a standard deviation of 0.5. And each coherence value persisted for a short interval. The interval duration was sampled from an exponential distribution with a mean of 270 milliseconds and then truncated to the range of 10 milliseconds to 1000 milliseconds. Additionally, the experiment comprised four blocks (each 5 minutes in duration): two under a 'frequent' condition and two under a 'rare' condition. The two conditions differed exclusively in the duration of intervals in which the signal coherence level was zero. Specifically, zero-signal-coherence intervals lasted 3-8 seconds in the frequent condition but were significantly longer (5-40 seconds) in the rare condition. Consequently, periods of sustained directional motion (non-zero signal coherence level) occurred more frequently in the frequent condition. All parameters governing coherence sequence generation and condition specifications were consistent with prior research[23].

The signal coherence level dictated the task phase. When a non-zero coherence period began (indicating leftward or rightward dot motion that was sustained for a period), patients were required to indicate the perceived direction of this sustained motion by pressing the corresponding button as quickly as possible. Consistent with previous research[23], for the operational definition, a response was considered correct only if made within a specific response window: starting 0.5 seconds after the non-zero period began and ending 0.5 seconds after it concluded. Responses made within this window but indicating the wrong direction were recorded as incorrect. Any participant response (correct or incorrect) within the response window immediately terminated that window and triggered the signal coherence level to revert to zero. False alarms occurred if participants responded outside the designated response windows described above. Following every response, participants received immediate visual feedback via the color of the central fixation dot: a green dot for a correct response, a yellow dot for an incorrect response, a red dot for a false alarm and a blue dot for miss (no response during the response window).

Prior to the formal experiment, participants completed a structured training protocol to ensure comprehension of the task rules, beginning with a verbal explanation of the cRDM task. Subsequently, they completed the first training block (guided practice) where they performed the task directly using high signal coherence levels ($\pm 0.7$); crucially, in addition to the standard central fixation dot feedback, this block provided explicit textual feedback (e.g., "Correct", "Incorrect") displayed after each response detailing performance. Advancement required achieving a correct response rate above 70%, after which participants proceeded to a second training block (observation and comprehension check) involving a demonstration video that explicitly illustrated the four distinct response outcomes (correct, incorrect, miss, false alarm) and their corresponding visual feedback colors. Following the video, participants were required to verbally describe each scenario and its associated feedback to the researchers to confirm understanding. Once comprehension was verified, participants entered the final training block (unguided practice), again performing the task with high coherence ($\pm 0.7$) but receiving only the standard central dot feedback, without textual descriptions. This unguided practice block terminated automatically once participants achieved a correct response rate above 70%.

**cRDM behavioral analysis**

Correct response rate was calculated as the proportion of correct responses within the response window. It was computed separately for each signal coherence level (|coherence| = 0.3, 0.4, 0.5) and for the frequent and rare conditions, and an overall mean was obtained by averaging across all signal coherence levels and conditions. False alarm rate was calculated for each condition as the total number of false alarms divided by the total time spent in that condition (10 minutes per condition), and an overall mean was obtained by averaging across conditions. In Study 1, we performed partial Spearman's correlations to examine the relationships between these behavioral measures and brain fog severity, controlling for age, sex, and education. In Study 3, the effect of TUS on cRDM performance was assessed with separate 2 × 2 repeated-measures ANOVAs for correct response rate and false alarm rate, with protocol (real vs sham) and time (baseline vs post-stimulation) as factors.

**Integration kernels and exponential decay model**

To characterize the pattern of motion coherence preceding responses, we calculated Integration Kernels. First, we extracted coherence time series in the 6-second window leading up to each response. To align the direction of evidence, we inverted the sign of the coherence values for all chunks ending in a "left" response. We then computed the Integration Kernels by separately averaging these processed time series for correct responses and false alarms. This analysis revealed a characteristic pattern: the Integration Kernels started near baseline, ramped up sharply, and peaked approximately 0.5 seconds before the response. Notably, the kernels for false alarms exhibited a steeper rise and a higher peak magnitude compared to those for correct responses.

To quantify this dynamic rise, we identified the peak time for each Kernels. We then truncated the Kernels at this peak and fitted an exponential model to the rising portion of the curve to describe its shape:

$$k(t) = Ae^{\frac{t}{T}}$$

In this model, $k(t)$ is Kernels value at time $t$, $A$ represents the amplitude (peak magnitude), and $T$ quantifies the steepness of the rise (a smaller $T$ indicates a faster increase). To estimate the parameters for each condition, we fitted the model by minimizing the sum of squared errors using MATLAB's fminsearch function. Finally, we used paired *t*-tests to compare the $A$ and $T$ values between correct responses and false alarms.

**GLM for evidence monitoring during cRDM**

Previous research indicates that perceptual decision-making relies on the temporally accumulated sensory evidence[43]. Based on our preceding analysis, we hypothesized that brain fog compromises the monitoring of steadily accumulated sensory evidence, leading to a greater reliance on fleeting spikes.

To test this hypothesis, we developed a General Linear Model (GLM) of the evidence-based decision-making process during the cRDM task. First, we processed the continuous coherence data to identify potential decision-making moments. We used a 100 ms sliding window, if the absolute average coherence within a window exceeded 0.3 (our defined lowest threshold for perceptible motion), we marked the window's end time as a candidate decision point.

Next, we quantified the magnitude of sensory evidence leading up to each decision, we defined two sets of regressors based on the absolute value of cumulative coherence within different time bins. This was done across two timescales within the 1.5 s preceding the candidate decision point, consisting of a "long-term evidence" set with three regressors from successive 0.5-second intervals, and a "short-term evidence" set with fifteen regressors from successive 0.1-second intervals. These two sets of regressors were then included in a single GLM. We then used a sigmoid function to map the computed value derived from these two sets of sensory evidence regressors to the probability of selecting "moving" (as opposed to "still").

Finally, we linked these regressors to the participant's actual behavior. A candidate point was labeled as "moving" (1) if the participant responded within the following 500 ms, and "still" (0) otherwise. We then separated these events to create distinct models for false alarms and correct responses, allowing us to test our primary hypothesis.

**Cognitive analysis in UKB**

The selection of cognitive metrics for our analysis was based on previous research from UKB that identified tasks most sensitive to cognitive impairment[11]. These metrics were derived from a data-driven approach that compared 778 UKB participants at risk for dementia with 778 matched controls. The cognitive assessment included the Symbol Digit Substitution test, TMT, Fluid Intelligence, Pairs Matching task, Card game "snap", and Numeric Memory[11].

To assess longitudinal changes in cognitive performance attributed to SARS-CoV-2 infection, we calculated the difference between follow-up and baseline scores for each of these cognitive metrics in our derived study cohort. To standardize these changes, each difference score was divided by its corresponding baseline value. Independent samples t-tests were then employed to compare the standardized change scores between the case and control groups for each cognitive metric.

**TUS protocol and simulation**

Ultrasound stimulation was delivered using NeuroFUS Transducer Power Output ™ (TPO-203) and H-104MR_4AA-4CH transducer (Sonic Concepts, Woodinville, WA, USA). For TUS group, we employed theta-burst TUS protocol[28,29,40] which consisted of stimulation in pulses of 20 ms within a pulse-repetition interval 200 ms (5 Hz pulse-repetition frequency; 10% duty cycle) and in total 400 cycles (80 seconds). The free-field spatial peak pulse-average intensity (ISPPA) was maintained at 33.8 W/cm$^2$.

Based on Study 1, the stimulation target region was defined as the overlapping region between the source localization results and the right inferior insula in the Destrieux atlas[41], with central MNI coordinates at (39, -16.5, 2.5). A pseudo-CT volume was then generated with T1-weighted images using a deep learning method[27]. Individual tissue compartments were reconstructed via simNIBS[67]. BabelBrain simulations[68] were used to optimize the ultrasound beam trajectory and focal depth to maximize precision.

Acoustic simulations confirmed that exposure remained within safety guidelines[69,70], and thermal simulations verified that the maximum target temperature increase did not exceed 2 °C. We used transducer placement optimization to maximize target ISPPA (7.370 ± 0.959 W/cm$^2$) and minimize off-target energy, while ensuring mechanical (MI = 0.693 ± 0.045) and thermal (CEM43 = 0.000384 ± 0.0000075) safety. Target ISPPA showed a 13% coefficient variation, aligning with inter-individual ranges of in-situ intensity in TUS studies[40,71] and indicating that variability between participants was controlled (See in **Extended Data Table 1**).

To ensure optimal contact, a custom gel pad matching the transducer's geometry was used to compensate for misalignment between the target stimulation site and the scalp's surface. Degassed ultrasound coupling gel (Aquasonic 100, Parker Laboratories Inc., Fairfield, NJ) was applied at the transducer-gel pad and pad-scalp interfaces for stable transmission. Before use, the gel was centrifuged to remove air bubbles, ensuring stable acoustic coupling. The ultrasound beam's angle of incidence on the skull was dictated primarily by the participant's unique head geometry.

Simulation-derived trajectories were imported into Brainsight version 2.5.4 (Rogue Research Inc., Montréal, Canada) for navigated guidance, where transducer and head positions were tracked in real-time at 1 Hz, maintaining targeting errors below 3 mm.

For sham stimulation, the ultrasound was inactive, but all procedural and auditory cues were maintained to mimic the active condition. The auditory stimuli consisted of the transducer's operational sound and a pre-recorded TPO sound. The transducer sound was delivered via bone-conduction headphones placed approximately 2 cm posterior to the temporal region, while the TPO sound was played through stereo speakers.

**Statistical analysis**

We presented the mean and standard error throughout both the figures and the manuscript. Statistical significance was reported using two-tailed *p*-values. For independent-sample *t*-tests, Cohen's *d* effect sizes were calculated as the difference between group means divided by the pooled standard deviation. For repeated measures ANOVA, effect sizes were reported as $\eta^2$, which represents the proportion of variance explained by the effect. Given the ordinal nature of the brain fog scale[30], and to ensure robustness against potential outliers, we used partial Spearman's correlations throughout.

**EEG acquisition and pre-processing**

EEG data was acquired using the Electrical Geodesics, Inc (EGI) system with a sampling rate of 1000Hz following HydroCel Geodesic Sensor Net 128-channel layout. All electrodes were referenced to Cz at the initial recording session. We used EEGLAB[61] to preprocess EEG data. First, we re-referenced the data to the average of left and right mastoids. Subsequently, a bandpass filter from 0.5Hz to 30Hz was applied to remove low frequency drifts and high frequency noise. Following this, we downsampled the data to 144 Hz to match the coherence series frequency of cRDM. Next, independent component analysis (ICA) was performed to identify and remove artifact-related components such as those associated with eye movements or muscle activity.

**Deconvolutional GLM analysis**

The deconvolutional GLM is often used in situations where neural activity evoked by stimulus presentations may exhibit overlap[34]. After matching EEG and coherence frequencies, we aligned EEG signal with events and time-expanded the design matrix with 11 regressors:

$$\begin{aligned} EEG\ Signal = &\ Correct\ Response + False\ Alarm + Coherence\ Change \\ &+ Response\ Period\ Onset + Response\ Period\ Onset_{|coherence|} \\ &+ Correct\ Rsponse_{signed} + False\ Alarm_{signed} + Coherence_{signed} \\ &+ |Coherence| + |Coherence\ Change| + Coherence\ Change_{difference} \end{aligned}$$

$Correct\ Response$, $False\ Alarm$, $Coherence\ Change$ and $Response\ Period\ Onset$ are stick functions (1 at event timepoints, 0 otherwise). $Correct\ Rsponse$ and $False\ Alarm$ unfold a time window from 5 seconds before to 3.5 seconds after. $Coherence\ Change$ unfolds a time window from 1 second before to 1.5 seconds after. $Response\ Period\ Onset$ represents the timepoint that a response period begins and unfolds a time window from 0.5 second before to 8 seconds after, while $Response\ Period\ Onset_{|coherence|}$ marks the same timepoint but includes signal coherence level information (0.3, 0.4, 0.5) and unfolds the same 0.5s-before-to-8s-after window.

$Correct\ Response_{signed}$ and $False\ Alarm_{signed}$ represent responses with directional information (-1 for left button presses, 1 for right button presses at the event timepoint), and unfold a time window from 5 seconds before to 3.5 seconds after.

$Coherence_{signed}$ and $|Coherence|$ represent the raw and absolute coherence series, respectively; $|Coherence\ Change|$ and $Coherence\ Change_{difference}$ represent the absolute and raw difference in coherence values before and after the coherence change,

respectively. All four regressors unfold a time window from 1.5 seconds before to 1.5 seconds after. We then solved the deconvolutional GLM to estimate parameters for these regressors at each channel for every subject, using the pseudoinverse method.

**Mediation analysis**

We employed the product of coefficients method with bootstrap resampling to investigate the mediating effect of the mediator PCPP magnitude (averaged over the time window from 0.4 to 0.6s after false alarm) in the relationship between the independent variable brain fog and the dependent variable false alarm. This approach involves estimating the total effect of brain fog on false alarm (path c), the direct effect of brain fog on false alarm while controlling for PCPP magnitude (path c'), the effect of brain fog on the mediator PCPP magnitude (path a), and the effect of PCPP magnitude on false alarm while controlling for brain fog (path b), with the indirect effect quantified as the product of paths a and b ($a \times b$). Statistical significance of the indirect effect was evaluated using 5,000 bootstrap resamples to generate 95% confidence intervals (CIs), where a significant mediation effect was indicated if the CI for $a \times b$ did not include zero.

**EEG source localization**

Source localization was performed in Brainstorm 3.241008[72], SPM12 (Statistical Parametric Mapping, Welcome Trust Centre for Neuroimaging, London, UK) and nilearn[73]. Using individual T1-weighted structural MRI data, we first segmented scalp, skull, and cortical gray matter boundaries via SPM12 to construct a three-compartment Boundary Element Method (BEM) head model[74]. And we solved the inverse problem using the standardized Low-Resolution Brain Electromagnetic Tomography Analysis (sLORETA) method[35]. Then, we epoched and averaged the localization data from 0.4 to 0.6 seconds after the false alarm response to investigate the main effect of PCPP. Second-level analysis was performed using a one-sample t-test implemented in nilearn to identify brain regions showing significant activation relative to baseline. Images were spatially normalized to MNI152 standard space and smoothed with an 8mm full-width at half-maximum (FWHM) Gaussian kernel. Multiple comparison correction was performed using a voxel-wise FWE correction based on the Bonferroni method, with a significance level of p < 0.001.

**MRI data acquisition (Study 1/ Study 3)**

Structural and resting-state magnetic resonance imaging (MRI) data were acquired using 3T scanners (Magnetom Spectra; Siemens Healthineers, Erlangen, Germany). For structural MRI, a 3D T1-weighted image was obtained with a magnetization-prepared rapid gradient-echo (MPRAGE) sequence, with parameters as follows: repetition time (TR) = 1690 ms, echo time (TE) = 2.4 ms, flip angle (FA) = 12°, matrix size = 256 × 256, slice thickness = 1 mm, and 200 sagittal slices. Additionally, in Study 3, 8 minutes of resting-state MRI data were collected using an echo planar imaging (EPI) sequence with the following parameters: 32 slices in an interleaved order, TR = 2000 ms, TE = 30 ms, FA = 90°, matrix size = 64 × 64, slice thickness = 3.5 mm, a 30° slice angle relative to the anterior-posterior commissure line was used to minimize signal loss in the insula cortex region.

**SBM analysis (Study 1)**

For SBM analysis, The T1-weighted (T1w) structural MRI data in Study 1 was preprocessed using the CAT12 v12.0 toolbox (integrated with SPM12 in MATLAB R2022b). Tissue segmentation and spatial normalization were performed via the CAT12 Segment function, with the MNI152NLin2009cAsym template[75] as the standard space. Quality control was implemented through visual inspection of gray/white matter segmentation results (using CAT12's Slice Display tool) and quantitative assessment via automated reports.

Preprocessed T1-weighted images were further processed to extract cortical thickness metrics. Using CAT12's Surface Tools module, cortical thickness maps were generated with input from central surface files derived from the segmentation step. This yielded vertex-wise thickness data for bilateral hemispheres, which were then smoothed using the Resample and Smooth Surf module with a 15-mm full-width at half-maximum (FWHM)

Gaussian kernel to enhance signal consistency. To derive regional cortical thickness metrics, smoothed thickness maps were parcellated using the Destrieux anatomical atlas[41]. Regional thickness values were aggregated to compute mean thickness for each Destrieux-defined region. To investigate whether cortical thickness correlates with the false alarm rate, we performed Spearman's correlations between the false alarm rate and mean cortical thickness in three regions of interest (ROIs) identified from PCPP source localization: the right inferior insula, left anterior middle cingulate cortex, and left posterior middle cingulate cortex.

**Structural MRI data analysis in UKB (Study 2)**

The Imaging-derived phenotype data used in Study 2 was generated by support team on behalf of UK Biobank[76], and made available to all researchers by UK Biobank[77]. Following the identification of longitudinal differences in perceptual ability attributed to SARS-CoV-2 infection, we aimed to investigate which longitudinal changes in brain regions within the case group predict these corresponding differences in perceptual ability. In line with our previous analysis, we used regional cortical thickness to predict longitudinal differences in perceptual ability. Due to cerebral symmetry, including bilateral regions (left and right hemispheres) in a single linear model could lead to unstable parameter estimates owing to potential collinearity between homologous regions of the two hemispheres. We therefore constructed separate GLM for the left and right hemispheres. Cortical thickness values for distinct brain regions, as defined by the Destrieux anatomical atlas[41] (consistent with Studies 1 and 2), were extracted from the UKB dataset. For each GLM, the regional cortical thicknesses were used to assess their specific predictive effects on longitudinal differences in perceptual ability.

**MRI data preprocessing (Study 3)**

Results included in Study 3 come from preprocessing performed using fMRIPrep 25.1.3[78] (RRID:SCR_016216), which is based on Nipype 1.10.0[79] (RRID:SCR_002502).

**Anatomical data preprocessing.** The T1w image was corrected for intensity non-uniformity with N4BiasFieldCorrection[80], distributed with ANTs 2.6.2[81] (RRID:SCR_004757), and used as T1w-reference throughout the workflow. The T1w-reference was then skull-stripped with a Nipype implementation of the antsBrainExtraction.sh workflow (from ANTs), using OASIS30ANTs as target template. Brain tissue segmentation of cerebrospinal fluid (CSF), white-matter (WM) and gray-matter (GM) was performed on the brain-extracted T1w using fast[82] (FSL 6.0, RRID:SCR_002823). Volume-based spatial normalization to one standard space (MNI152NLin2009cAsym) was performed through nonlinear registration with antsRegistration (ANTs 2.6.2), using brain-extracted versions of both T1w reference and the T1w template. The following template was selected for spatial normalization: ICBM 152 Nonlinear Asymmetrical template version 2009c[75] (RRID:SCR_008796; TemplateFlow ID: MNI152NLin2009cAsym).

**Functional data preprocessing.** For each of the BOLD runs per subject (across all sessions), the following preprocessing was performed. First, a reference volume and its skull-stripped version were generated using a custom methodology of fMRIPrep. A B0 nonuniformity map (or fieldmap) was estimated based on a phase-drift map calculated with two consecutive GRE (gradient-recalled echo) acquisitions. The corresponding phase-map(s) were phase-unwrapped with prelude (FSL 6.0). The fieldmap was then co-registered to the target EPI (echo-planar imaging) reference run and converted to a displacements field map (amenable to registration tools such as ANTs) with FSL's fugue and other SDCFlows tools. Based on the estimated susceptibility distortion, a corrected EPI (echo-planar imaging) reference was calculated for a more accurate co-registration with the anatomical reference.

The BOLD reference was then co-registered to the T1w reference using flirt[83] (FSL 6.0) with the boundary-based registration[84] cost function. Co-registration was configured with six degrees of freedom to account for distortions remaining in the BOLD reference. Head-motion parameters with respect to the BOLD reference (transformation matrices, and six corresponding rotation and translation parameters) were estimated before any spatiotemporal filtering using mcflirt[83] (FSL 6.0).

Several confounding time-series were calculated based on the preprocessed BOLD: framewise displacement (FD), DVARS and three region-wise global signals. FD was computed using two formulations following Power[85] (absolute sum of relative motions) and Jenkinson[83] (relative root mean square displacement between affines). FD and DVARS were calculated for each functional run, both using their implementations in Nipype (following the definitions by Power et al.[85]). The three global signals were extracted within the CSF, the WM, and the whole-brain masks. Additionally, a set of physiological regressors were extracted to allow component-based noise correction[86] (CompCor). Principal components were estimated after high-pass filtering the preprocessed BOLD time-series (using a discrete cosine filter with 128s cut-off) for the two CompCor variants: temporal (tCompCor) and anatomical (aCompCor). tCompCor components were then calculated from the top 2% variable voxels within the brain mask. For aCompCor, three probabilistic masks (CSF, WM and combined CSF+WM) were generated in anatomical space. The implementation differed from that of Behzadi et al.[86] instead of eroding the masks by 2 pixels on BOLD space, the aCompCor masks were subtracted by a mask of pixels that likely contain a volume fraction of GM. This mask is obtained by thresholding the corresponding partial volume map at 0.05, and it ensured components were not extracted from voxels containing a minimal fraction of GM. Finally, these masks were resampled into BOLD space and binarized by thresholding at 0.99 (as in the original implementation). Components were also calculated separately within the WM and CSF masks. For each CompCor decomposition, the k components with the largest singular values were retained, such that the retained components' time series is sufficient to explain 50 percent of variance across the nuisance mask (CSF, WM, combined, or temporal). The remaining components were dropped from consideration. The head-motion estimates calculated in the correction step were also placed within the corresponding confounds file. The confound time series derived from head motion estimates and global signals were expanded with the inclusion of temporal derivatives and quadratic terms for each[87]. Frames that exceeded a threshold of 0.5 mm FD or 1.5 standardized DVARS were annotated as motion outliers.

Additional nuisance time series were calculated by means of principal components analysis of the signal found within a thin band (crown) of voxels around the edge of the brain, as proposed by Patriat et al.[88]. All resamplings can be performed with a single interpolation step by composing all the pertinent transformations (i.e. head-motion transform matrices, susceptibility distortion correction when available, and co-registrations to anatomical and output spaces). Gridded (volumetric) resamplings were performed using nitransforms, configured with cubic B-spline interpolation.

**Resting-state Functional Connectivity Analysis (Study 3)**

Following preprocessing, FC analysis was conducted using Python scripts based on the nilearn library[73]. The cerebral cortex was parcellated into 148 anatomical regions according to the Destrieux atlas[41], with the 5 mm-radius spherical seed[89,90] in the right inferior insula included as the seed region (Extended Data Fig. 6a). Subjects exhibiting excessive head motion—defined as a mean FD greater than 0.3—were excluded from further analysis. For each remaining subject, nuisance signals were addressed as follows: the first five volumes were discarded to ensure T1 equilibrium, and confound regression was performed to remove the mean signals from white matter, cerebrospinal fluid, and the global signal. High-motion volumes were identified and censored using a scrubbing strategy, applying thresholds of $FD > 0.5$ and standardized $DVARS > 1.5$. After denoising, the mean time series for each region was extracted from the cleaned functional data, which was further detrended, high-pass filtered at 0.01 Hz, and spatially smoothed with a 3 mm FWHM kernel. FC coefficients were then generated by computing the Pearson correlation between the time series of the stimulation target and all other regions (excluding the right inferior insula in the Destrieux atlas[41], as it overlapped with the stimulation target). Finally, the resulting correlation coefficients were Fisher-Z transformed to normalize their distribution for subsequent statistical analyses.

To statistically assess the effects of the TUS protocol and inter-regional distance on changes in FC, we constructed an LME model. The analysis was performed in R (Version 4.4.1) using lme4 package. The change in FC from baseline- to post-stimulation was entered as the dependent variable. The model's fixed effects included the main effect of the stimulation Protocol (TUS/Sham stimulation) and controlled the effects of distance in a protocol-specific manner. Given that different brain regions exhibit distinct intrinsic properties and connectivity profiles, we anticipated variability in their response magnitudes. So, the model included a random-effects structure of (1 | ROI), which models random intercepts for ROIs. The formula is as follows:

$$FC_{change} \sim Protocol + Protocol:Distance + (1|ROI)$$

Group comparisons on Protocol effects were subsequently conducted using estimated marginal means (emmean) in R.

**Extended Data Figures**

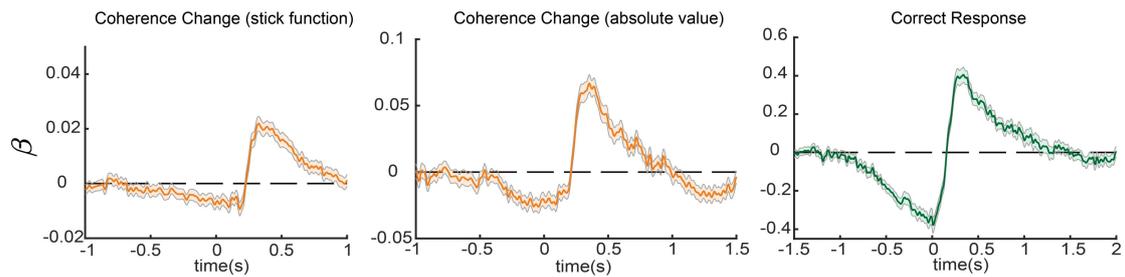

**Extended Data Fig. 1 | Neural activity in the centroparietal region for other notable regressors.**

The neural response time-locked to the onset of a change in stimulus coherence, modeled as an impulse (stick function), shows a significant positive-going potential after the event. This waveform is characteristic of the Centroparietal Positivity (CPP), often linked to evidence accumulation in perceptual decision-making process (Left). A similar CPP component was observed for a parametric regressor representing the absolute magnitude of the coherence change, indicating sensitivity to the strength of the updated evidence (Middle). For correct responses, the neural activity exhibited a distinct biphasic pattern: a significant negative deflection (deactivation) immediately preceding the motor response, followed by a large and rapid positivity. This post-response positivity distinguishes correct trials from false alarms (Right).

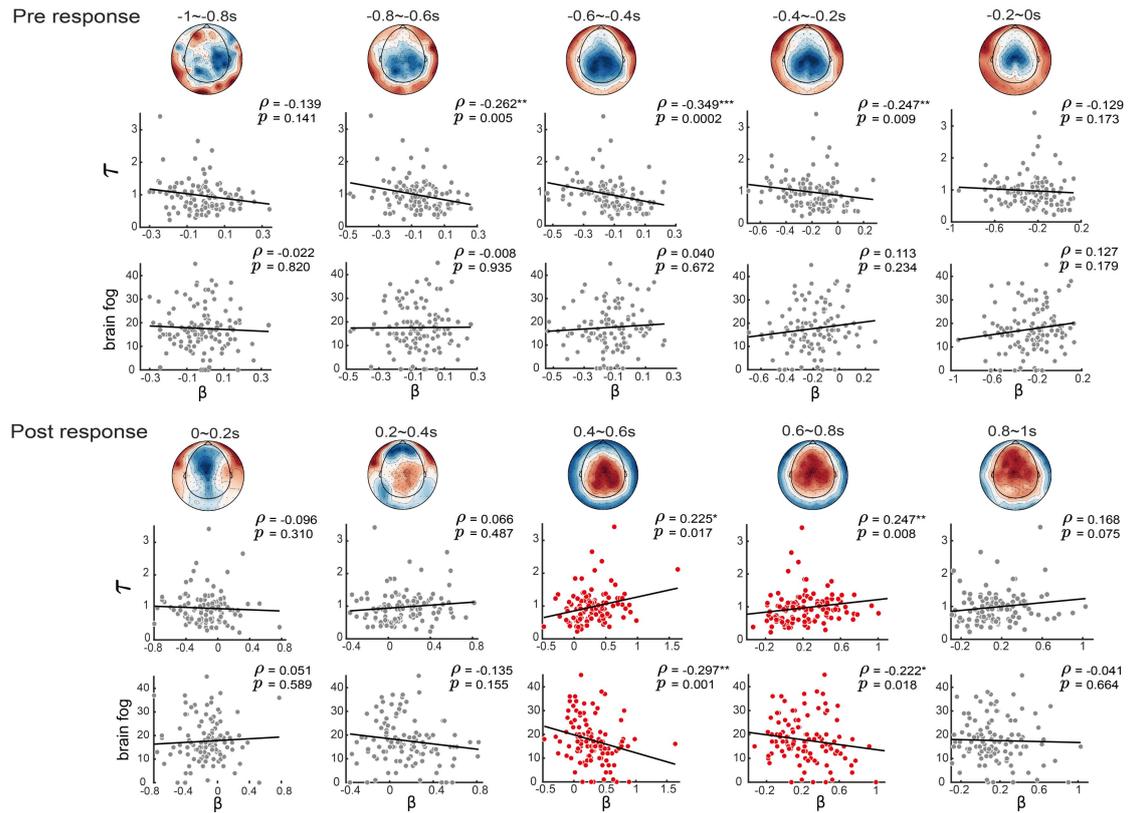

**Extended Data Fig. 2 | Neural activity during false alarms correlates with T and brain fog.**

To identify the optimal time window for our analysis, we correlated neural activity with $T$ and brain fog within the 2-second period surrounding a false alarm (-1 to +1 s), using a 0.2 s sliding window. The results show that significant correlations with both $T$ and brain fog were confined to the 0.4–0.6 s and 0.6–0.8 s post-alarm time windows (red). Consequently, we selected the 0.4 – 0.6 s window for all subsequent analyses due to its stronger correlation with brain fog.

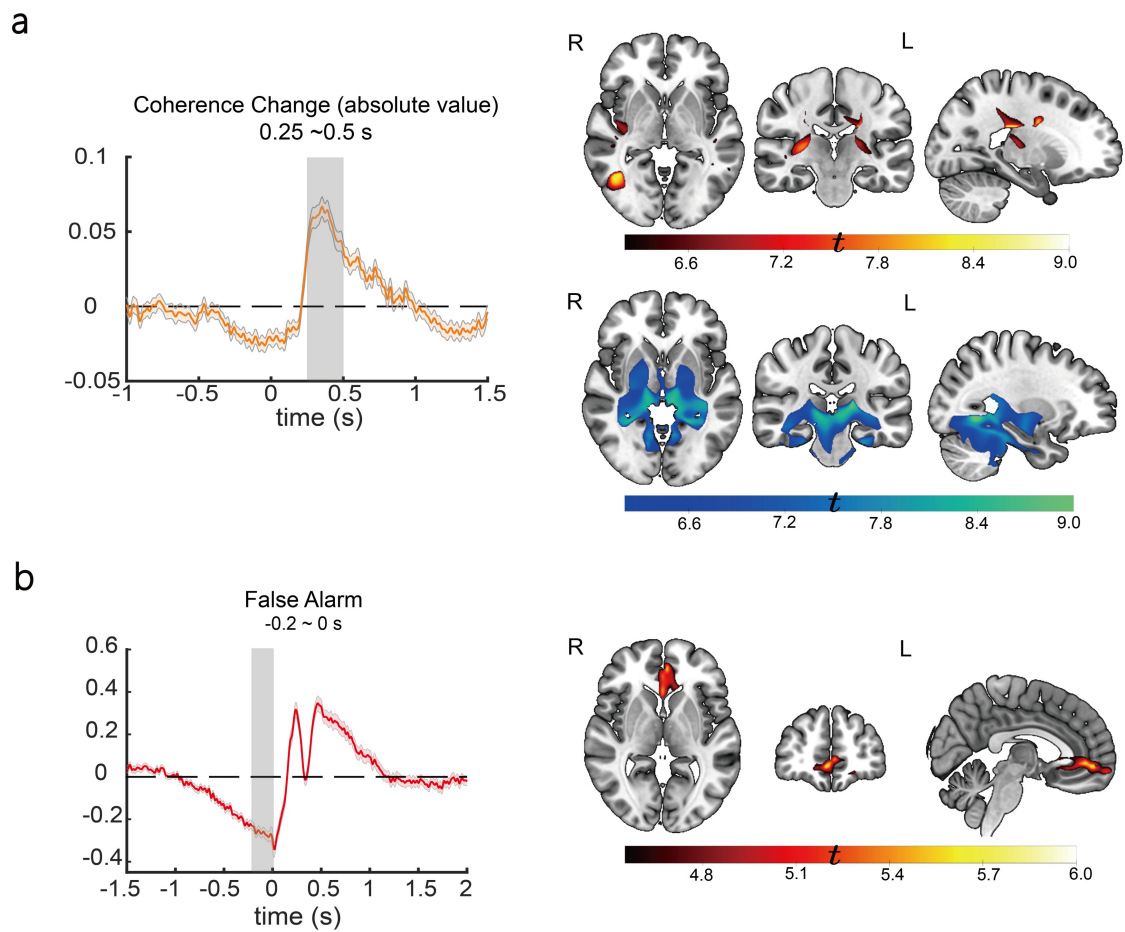

**Extended Data Fig. 3 | Source localization of representative electrophysiological potentials during perceptual processing.**

**a,** Source localization of the Centroparietal Positivity (CPP), an electrophysiological marker of evidence accumulation[23]. The analysis revealed significant activation in the right anterior occipital cortex, inferior insula and middle cingulate cortex, with concurrent deactivation in limbic and parahippocampal regions (voxel-wise FWE corrected, based on Bonferroni method, $p < 0.001$).

**b,** Source localization of the Readiness Potential (RP), a neural correlate of movement preparation[91]. This analysis revealed activation in the orbitofrontal and anterior cingulate cortices (voxel-wise FDR corrected, $p < 0.001$; No significant results for Bonferroni correction).

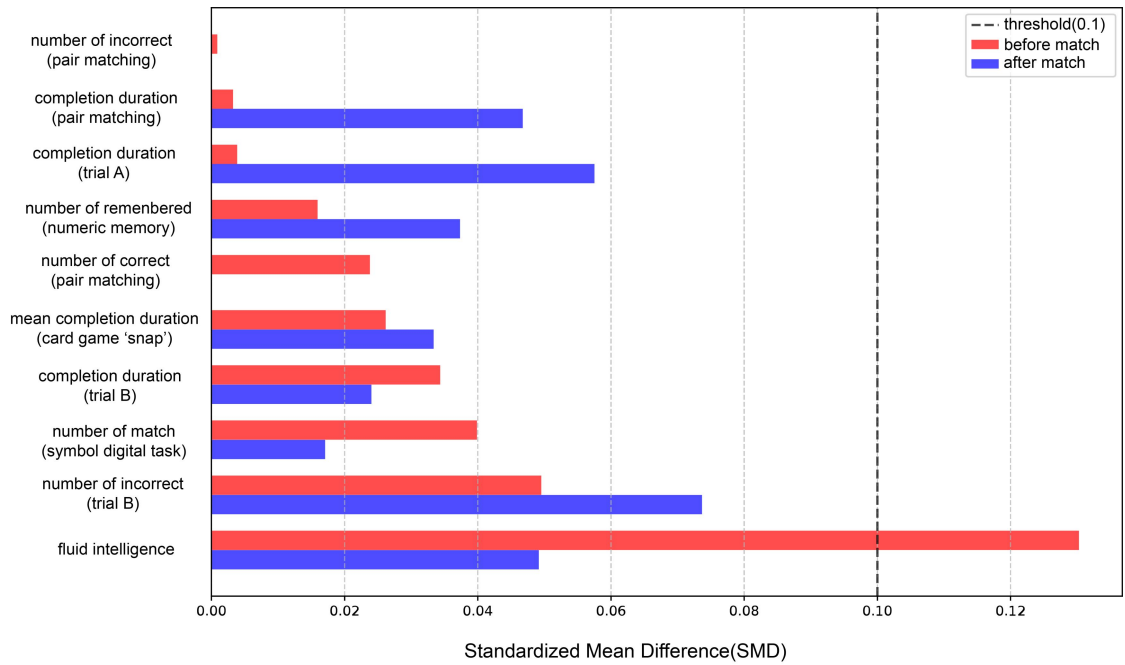

**Extended Data Fig. 4 | Standardized Mean Difference (SMD) for cognitive measures before and after Mahalanobis distance matching.**

The figure displays the SMD for ten cognitive measures comparing case and sham groups within the UKB. Prior to matching, a significant imbalance was evident, most notably in fluid intelligence, which had an SMD greater than the 0.1 threshold (red bars). To create well-matched groups while accounting for potential correlations between cognitive domains, we applied Mahalanobis distance matching. Following this procedure, all ten measures were balanced, as indicated by all SMDs being well below the 0.1 threshold (blue bars).

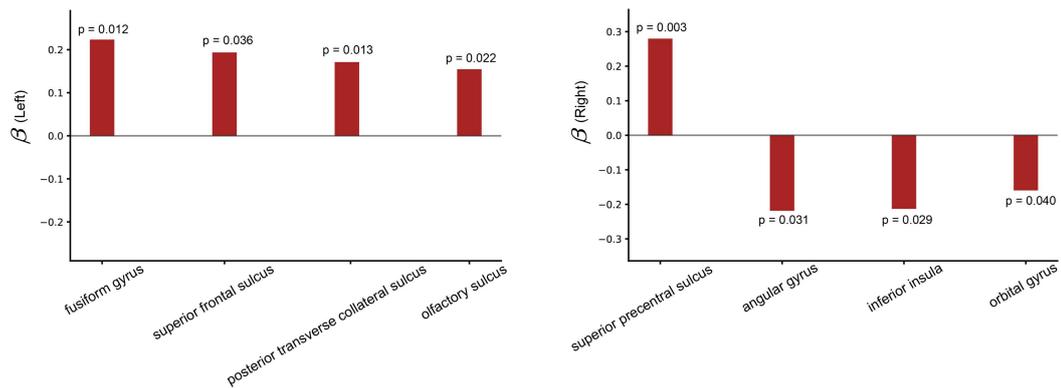

**Extended Data Fig. 5 | Brain regions where cortical thickness changes significantly predict changes in TMT-A performance.**

Results from GLM analysis identifying significant associations between changes in regional cortical thickness and changes in TMT-A completion time. The bars represent the standardized beta ($\beta$) coefficients for each significant region, displayed for the left and right hemispheres. A positive $\beta$ indicates that decreased thickness was associated with shorter completion times (better performance), while a negative $\beta$ indicates decreased thickness was associated with longer completion times (poor performance).

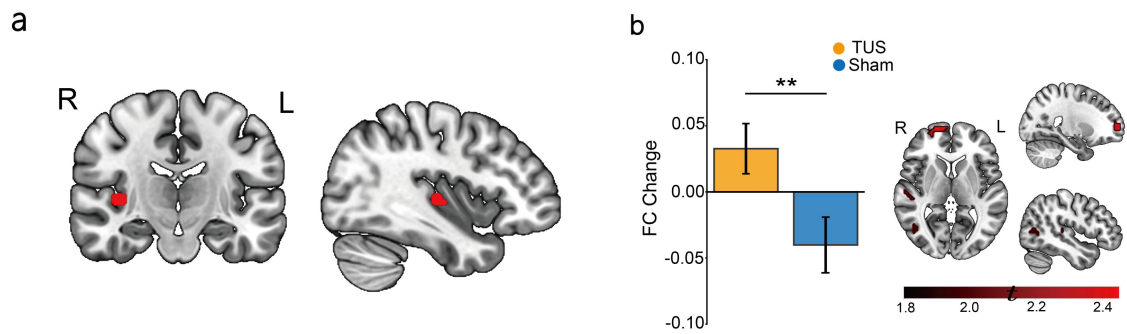

**Extended Data Fig. 6 | TUS Enhances Functional Connectivity of the Right Insula.**

**a,** Insula region of interest. A 5 mm-radius spherical region of interest (red) in the right inferior insula was used as the seed for connectivity analysis (centered on the TUS target region's central coordinate).

**b,** FC changes. LME analysis indicated that TUS induced positive changes in insula-centered FC compared to Sham stimulation. The brain map (right) shows regions with increased connectivity from the insula seed in the TUS vs. Sham group (*t*-statistic map, *p* < 0.05 uncorrected). Connectivity gains were observed in the right frontal pole, right superior temporal gyrus, and right anterior occipital cortex. These neural changes accompanied the behavioral improvement in the TUS group.

**Extended Data Table**

**Extended Data Table 1 | BabelBrain transcranial simulations**

| ID | Focal Depth (mm) | Target $I_{SPPA}$ (W/cm$^2$) | MI | Target Temp$_{max}$ (℃) | Target CEM43 |
|---|---|---|---|---|---|
| 1 | 61 | 5.968 | 0.625 | 37.159 | 0.00038 |
| 2 | 52 | 7.926 | 0.721 | 37.175 | 0.00039 |
| 3 | 51 | 7.950 | 0.721 | 37.168 | 0.00038 |
| 4 | 49 | 6.439 | 0.650 | 37.135 | 0.00038 |
| 5 | 47 | 7.930 | 0.721 | 37.161 | 0.00039 |
| 6 | 48 | 6.955 | 0.675 | 37.143 | 0.00038 |
| 7 | 47 | 5.958 | 0.625 | 37.129 | 0.00037 |
| 8 | 53 | 5.965 | 0.625 | 37.137 | 0.00038 |
| 9 | 46 | 6.928 | 0.675 | 37.141 | 0.00038 |
| 10 | 47 | 5.958 | 0.625 | 37.129 | 0.00037 |
| 11 | 49 | 7.452 | 0.698 | 37.168 | 0.00039 |
| 12 | 48 | 7.941 | 0.721 | 37.167 | 0.00039 |
| 13 | 49 | 7.941 | 0.721 | 37.164 | 0.00039 |
| 14 | 47 | 7.940 | 0.721 | 37.156 | 0.00038 |
| 15 | 50 | 9.453 | 0.786 | 37.189 | 0.00040 |
| 16 | 47 | 7.930 | 0.721 | 37.161 | 0.00039 |
| 17 | 55 | 7.452 | 0.698 | 37.158 | 0.00038 |
| 18 | 51 | 6.920 | 0.675 | 37.150 | 0.00038 |
| 19 | 48 | 7.944 | 0.721 | 37.170 | 0.00039 |
| 20 | 62 | 8.451 | 0.744 | 37.175 | 0.00039 |

**Note.** For each participant in TUS group, the table lists focal depth, the target $I_{SPPA}$, and the global maximum values of MI, maximum temperature and CEM43 at the target location.

**Acknowledgment**

Conceptualization, Y.L., Y.T.; Investigation, J.Y., S.Z., Z.W., J.X., J.C., Z.Y., T.W., C.G., X.L., X.L., X.Z., and K.L.; Writing – Original Draft, J.Y., Y.L.; Writing – Review & Editing, J.Y., Y.L., Y.T., R.D., and R.G. We also thank Prof. Tianye Jia for providing critical comments on an earlier version of the manuscript.

This study is supported by the Noncommunicable Chronic Diseases-National Science and Technology Major Project (2025ZD0546200), the National Science and Technology Innovation 2030 Major Programme (2022ZD0205500), the National Natural Science Foundation of China (32271093), the Beijing Natural Science Foundation (Z230010, L222033), the Beijing Outstanding Young Scientist Program (JWZQ20240101023) and the Fundamental Research Funds for the Central Universities.

**Conflict of interest**

The authors have indicated they have no potential conflicts of interest to disclose.

**Data availability**

EEG, fMRI and behavioural data that support the conclusions in this study will be available upon publication. The raw individual participant imaging data will be available upon reasonable request to the corresponding author, subject to participant consent.

**Code availability**

The analysis code will be publicly available upon publication.